\documentclass[11pt, showpacs]{article}
\pdfoutput=1

\usepackage{amsmath,amssymb,amsfonts,bm,marginnote,cite,graphicx,slashed,bbm}
\usepackage[colorlinks=true, pdfstartview=FitV, linkcolor=black, citecolor=black, urlcolor=black]{hyperref}
\usepackage[usenames,dvipsnames]{xcolor}

\usepackage{epstopdf}
\usepackage[justification=centering]{caption}
\usepackage{subcaption}
\usepackage{dcolumn}
\usepackage{bbm}
\usepackage{amscd}
\usepackage{mathrsfs}
\usepackage{dsfont}
\usepackage{comment}
\usepackage{setspace}
\usepackage[margin=0.75in]{geometry}
\usepackage{tensor}
\usepackage{mathrsfs}
\usepackage{cancel}
\usepackage{hyperref}
\usepackage{empheq}
\usepackage{youngtab}
\usepackage{empheq}
\usepackage{MnSymbol}
\usepackage{float}
\usepackage[all]{xy}
\usepackage{graphbox}
\usepackage{longtable}
\usepackage{subcaption}

\DeclareMathAlphabet{\mathbbold}{U}{bbold}{m}{n}

\edef\marginnotetextwidth{\the\textwidth}

\parskip=\baselineskip

\newcommand{\thistitle}{
Knots, links, and long-range magic}
	
\newcommand{\addressam}{
	Institute of Physics, University of Amsterdam,
	904 Science Park, 1098 XH Amsterdam, The Netherlands
	}

\newcommand{\be}{\begin{equation}}
\newcommand{\ee}{\end{equation}}
\newcommand{\beq}{\begin{eqnarray}}
\newcommand{\eeq}{\end{eqnarray}}
\newcommand{\bea}{\begin{eqnarray}}
\newcommand{\eea}{\end{eqnarray}}
\newcommand{\beqn}{\begin{eqnarray}}
\newcommand{\eeqn}{\end{eqnarray}}

\newcommand{\bs}{\boldsymbol}

\def\pa{\partial}

\def\Tr{\text{Tr}}

\newcommand{\mTr}{\mathrm{Tr}}

\newcommand{\mf}[1]{\mathfrak{#1}}
\newcommand{\mc}[1]{\mathcal{#1}}

\newcommand{\tmc}[1]{\tilde{\mathcal{#1}}}

\begin{document}

\title{\thistitle}
\author{
	{Jackson R. Fliss$^{1}$}\\
	\\
	{$^1$\small \emph{\addressam}}
\\}
\date{\today}
\maketitle\thispagestyle{empty}

\begin{abstract}
We study the extent to which knot and link states (that is, states in 3d Chern-Simons theory prepared by path integration on knot and link complements) can or cannot be described by stabilizer states.  States which are not classical mixtures of stabilizer states are known as ``magic states" and play a key role in quantum resource theory.  By implementing a particular magic monotone known as the ``mana" we quantify the magic of knot and link states.  In particular, for $SU(2)_k$ Chern-Simons theory we show that knot and link states are generically magical.  For link states, we further investigate the mana associated to correlations between separate boundaries which characterizes the state's long-range magic.  Our numerical results suggest that the magic of a majority of link states is entirely long-range.  We make these statements sharper for torus links.

\end{abstract}

\newpage

\section{Introduction}\label{sect:Intro}
Computational complexity has become the focus for a growing set of investigations into the structure of quantum field theories.  These investigations have also been fruitful in the realm of holography, providing insightful connections between semi-classical gravity and quantum information beyond the scope of entanglement \cite{Susskind:2014rva,Brown:2015bva,Goto:2018iay,Harlow:2013tf,Brown:2019rox}.  The key concept of these approaches is ``circuit complexity," defined by counting some number of distinguished ``simple gates" needed to build a state from a fixed fiducial state.  This approach has bootstrapped itself nicely from finite dimensional quantum systems to continuum quantum field theory \cite{Jefferson:2017sdb,Chapman:2017rqy,Khan_2018,Caceres:2019pgf} through Nielsen's geodesic formulation of circuit complexity\cite{Nielsen:aa,Dowling:aa} and through continuum tensor networks \cite{Caputa:2017urj,Caputa:2017yrh}.\\
\\
The distinction of ``simple" and ``hard" gates however begs the following question: what {\it types} of states can one build from simple gates alone?  Or: how many ``hard" gates does one need to build a generic state?  This suggests a complementary approach to computational complexity, which assigns complexity to the number of ``hard" gates needed to prepare a state,  or, alternatively, how many hard gates one can bypass by the injection of the state as a resource.  In finite dimensional quantum systems, quantifying and leveraging this notion of complexity is the goal of {\it quantum resource theory} (QRT)\cite{HORODECKI_2012, veitch2012negative,Veitch:2014aa}.  In this context, the simple gates stabilize a set of generalized Pauli operators and the simple states prepared by their circuits are {\it stabilizer states.}  This particular declaration of ``simple" gates is physically motivated: such circuits can be efficiently simulated by classical computation.  This also complements the circuit complexity mantra ``{\it entanglement is not enough}" \cite{Susskind:2014moa} as the ``simple states", i.e. the stabilizer states, can display both short-range and long-range entanglement.  In this alternative vantage point, one would declare a state complex if it is non-stabilizer and would like a quantifiable measure of its ``non-stabilizer-ness."  In QRT, such states are referred to as {\it magic states} and their magic is quantified by {\it magic monotones.}  The monotone of focus in this paper is known as the {\it mana} and we will define it shortly below.\\
\\
The generalization of these concepts to continuum quantum field theory is still preliminary, yet has already yielded interesting results with potential implications for conformal field theory and phase transitions \cite{White:2020zoz, Sarkar_2020} as well as holography\cite{Balasubramanian:2005mg, Balasubramanian:2008da}.  However, the resource theory of magic states is still a relatively open and unexplored topic in quantum field theory.  Even more mysterious is the subregion distribution of magic and its implications.  The aim of the current paper is to add another entry into this small but growing lexicon.  Our approach will be to ``split-the-difference" between quantum mechanics and quantum field theory and focus on states in {\it topological quantum field theory} (TQFT) \cite{Atiyah1988,Witten:1988ze}.  Because TQFTs typically have no local propagating degrees of freedom, the Hilbert spaces of these theories are determined solely by the topology on which the space is defined.  TQFTs strike a delicate balance of simplicity and universality (often appearing as effective field theories of IR topological phases), and so make excellent arenas for exploring concepts in quantum information at the boundary of quantum mechanics and quantum field theory.\\
\\
As a notable instance of this, in \cite{Salton:2016qpp, Balasubramanian:2016sro, Balasubramanian:2018por} (see also \cite{1711.06474,Dwivedi:2017rnj, Chun:2017hja}), the long-range multi-partite entanglement in Chern-Simons theory was investigated via states associated with links embedded in the three-sphere, $S^3$.  Naturally, such states were coined {\it link states.}  With ``{\it entanglement is not enough}" in mind, we extend this story to realm of magic, defining a notion of complexity to these states \footnote{We are also aware of forthcoming work on the conventual ``circuit complexity" of knot and link states: \cite{LeighPai:2020}.}.  Because the magic also characterizes pure states, we can also discuss states associated to knots (i.e. {\it knot states}) without having to trace out any components.  We will find that the mana reveals a certain universality to knot states that cannot be characterized by entanglement.  We additionally look at the mana of a subset of link states.  By tracing over the Hilbert space corresponding to a sub-component of a link, we can investigate the structure of {\it long-range magic}, that is magic that is distributed over arbitrary length scales.  We will find that the topology of a link can greatly determine the structure and the robustness of its long-range magic.\\
\\
The paper is organized as follows.  In section \ref{sect:setup} we review the construction of knot and link states in Chern-Simons theory.  We also provide a lightning overview of the necessary facts and definitions regarding magic and mana, including a basic review on Clifford unitaries, stabilizer states, and the discrete Wigner transform.  In section \ref{sect:AbelianCS} we characterize the (non)magic of all states in Abelian Chern-Simons theory and thus necessitate the study of non-Abelian Chern-Simons theory.  For the rest of the paper we focus on $SU(2)_k$ Chern-Simons theory at various levels, $k$. In section \ref{sect:knots} we present our characterization of the magic associated to knot states.  Then in section \ref{sect:LRM} we begin our investigation of the long-range magic associated to two-links.  We then discuss our results and our outlook in the broader context of QRT and quantum field theory in section \ref{sect:disc}.  Lastly, in the appendices we provide greater detail of the knot/link state construction (appendix \ref{app:braidcomp}), a tabulation of mana for a list of knot states (appendix \ref{app:listknots}), and a tabulation of the long-range mana for a list of two-links (appendix \ref{app:linktable}).

\section{Setup}\label{sect:setup}
\subsection{Knot and link states}\label{set:statedefs}
\begin{figure}[h!]
\centering
\begin{subfigure}[c]{.375\textwidth}
\centering
\includegraphics[width=\textwidth]{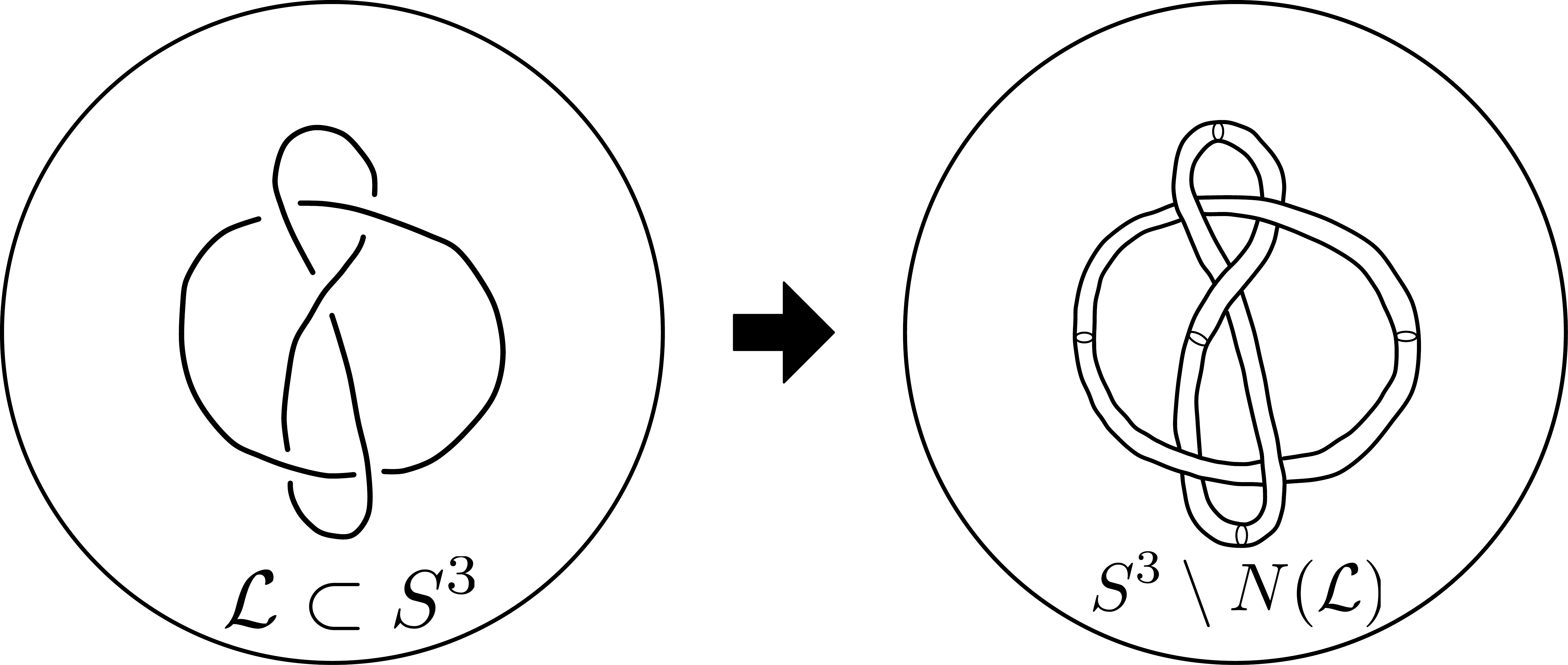}
\end{subfigure}
\quad,\quad
\begin{subfigure}[c]{.55\textwidth}
\centering
\includegraphics[width=\textwidth]{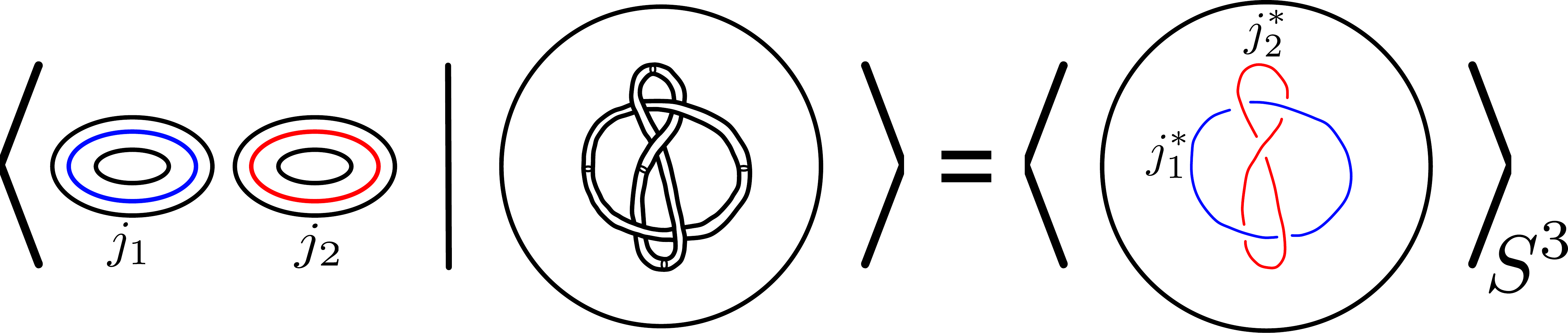}
\end{subfigure}
\caption{\small{\textsf{(Left) From a link $\mc L$, obtaining the link complement by deleting $N(\mc L)$ from $S^3$.  (Right) The wave-function of the associated state in the {\bf rep basis} is given by the $S^3$ expectation values of Wilson loops.}}}
\label{fig:WFdef}
\end{figure}

Let us briefly review the construction of the states in question (a more detailed construction can be found in \cite{Balasubramanian:2016sro}).  The key tool will be the Chern-Simons path integral over a 3-manifold $\mathscr M$ (to be described shortly below).  The action for this theory is
\beq\label{eq:CSaction}
S=\frac{k}{4\pi}\int_{\mathscr M} \mTr\left(A\wedge dA+\frac{2}{3}A\wedge A\wedge A\right)
\eeq
$A$ is a one-form connection on the principal $G$-bundle over $\mathscr M$, $k\in\mathbb Z$ is the level, and $\mTr$ is the Killing-form inner product on the associated Lie algebra, $\mf g$\footnote{For $SU(N)_k$  this can be taken to be the trace in the fundamental representation.}.  It is customary to label this theory as $G_k$ to emphasize the roles of both the group and the level.\\
\\
This theory is a well-known topological quantum field theory (albeit with a small caveat to be described in the next section) and as such, when $\mathscr M$ has a boundary, it associates a Hilbert space factor to each connected component of $\pa\mathscr M$.  The wave-functionals of the states furnishing that Hilbert space are produced by path-integration.  For the current purposes we are interested in states on torus boundaries associated to a link\footnote{We will frame this discussion in terms of links but the corresponding statements also apply to knots as ``1-component links".} in the following way.  Given an $n$-component link, $\mc L=L_1\cup L_2\cup\ldots \cup L_n$, we embed it as a set of non-intersecting circles into $S^3$.  We then imagine ``fattening" each circle, defining a set of non-intersecting tubular neighborhoods containing each component (we will call this set $N(\mc L)$) and then deleting that set from $S^3$: $\mathscr M=S^3\setminus N(\mc L)$.  Such a manifold is called the {\it link complement of $\mc L$ in $S^3$}.  The boundary of $\mathscr M$ is given by the disjoint union of tori, each associated to a particular component of the link:
\beq
\pa \mathscr M=\bigcup_{i=1}^n T^2
\eeq
the resulting Hilbert space, $\mc H_{\pa\mc M}$, is the tensor product over torus Hilbert spaces
\beq
\mc H_{\pa\mc M}=\bigotimes_{i}^n\mc H_{T^2}
\eeq
and the path-integral produces a unique state $|\mc L\rangle$ in that Hilbert space.  Let us describe this state in more detail.  A given $\mc H_{T^2}$ is spanned by a basis labelled by the integrable highest weight representations, $\{\mc R_j\}$, of the affine Lie algebra, $\mf g_k$.  For $\mf{su}(2)_k$, $j$ are half-integers ranging from $0$ to $k/2$.  One can imagine that one such basis state, $|j\rangle$, is prepared by the path-integration over the solid torus with a Wilson line in the $\mc R_j$ representation threading the interior longitude.  We will call this basis the {\bf representation basis} (or {\bf rep basis}) to distinguish it from a different choice of basis that will follow.\\
\\
Via the axioms of TQFT, the wave-function of the link state $|\mc L\rangle$ in the {\bf rep basis}
\beq
\Psi_{\mc L}(j_1,\ldots, j_n)=\langle j_1,j_2,\ldots, j_n|\mc L\rangle
\eeq
is given by the path-integral over the manifold resulting from gluing $S^3\setminus N(\mc L)$ to the solid tori preparing each $|j_i\rangle$ (with their orientation reversed) along their common boundary.  This has the effect of filling in each tubular neighborhood of $N(\mc L)$ along with a Wilson loop along each circle, $L_i$, in the conjugate representation $j_i^\ast$: $W_{j_i^\ast}(L_i)=\mTr_{\mc R_{j_i^\ast}}\mc P\exp\left(\oint_{L_i}A\right)$.  This is depicted as a cartoon in figure \ref{fig:WFdef}.  Thus the wave-function of the link state in this basis is precisely the $S^3$ expectation value of a set of knotted and linked Wilson loops, and these are precisely the {\it colored Jones polynomials} associated to the link (with a particular normalization) \cite{Witten:1988hf}:
\beq
\Psi_{\mc L}(j_1,j_2,\ldots, j_n)=\big\langle W_{j_1^\ast}(L_1)W_{j_2^\ast}(L_2)\ldots W_{j_n^\ast}(L_n)\big\rangle_{S^3}=J_{j_1^\ast,j_2^\ast,\ldots, j_n^\ast}(\mc L)
\eeq
\\
Let us take a brief moment to discuss the issue of {\it framing}.  In truth, the quantum path-integral is not a strict topological object {\it per se}; it also requires a choice of framing.  For a particular component, say $L_i$, of $\mc L$ this can be described by a choice of normal vector field to $L_i$; the framing then counts the winding of this vector field as it traverses the entire $L_i$ cycle.  This is equivalent to regulating the self-linking of the $L_i$ component.  In principle, the colored Jones polynomial depends on the choices of framing and this is the sole geometric data needed to specify the Wilson line expectation values.  Fortunately, the Chern-Simons path-integral responds in a simple way to changing the framing of $L_i$ by $n$ twists:
\beq
\Psi_{\mc L}(j_1,j_2,\ldots,j_i,\ldots, j_n)\rightarrow\left(\mc T_{j_i}\right)^n\Psi_{\mc L}(j_1,j_2,\ldots,j_i,\ldots,j_n)
\eeq
where $\mc T$ is the modular T-matrix (diagonal in the {\bf rep basis}):
\beq
\mc T_{j}=\exp\left(2\pi i h_j\right)
\eeq
and $h_j$ is the conformal weight associated to $j$.  For $\mf{su}(2)_k$ we have $h_j=\frac{j(j+1)}{k+2}$.  One notes that $\mc T$ clearly acts as a local unitary on our basis states.  Thus for many purposes (for instance investigating the entanglement structure of $|\mc L\rangle$) the framing can be safely ignored.  For the purposes of this paper, however, we will need to keep track of it, at least for now.\\
\\
Lastly, before moving onward to a review magic and mana, we will shortly introduce another set of basis states on $\mc H_{T^2}$ that we will use extensively in what follows.  We call this basis the {\bf Verlinde basis} and denote it by $\{|\tilde j\rangle\}$.  It is defined in terms of {\bf rep basis} as
\beq
|\tilde j\rangle=\sum_{j}{\mc S^\dagger_{\tilde j}}^j|j\rangle
\eeq
where $\mc S$ is the modular S-matrix which, along with $\mc T$, generates the $PSL(2,\mathbb Z)$ set of modular transformations on the torus:
\beq
\mc S:\;\tau\rightarrow -\frac{1}{\tau}\qquad\qquad \mc T:\;\tau\rightarrow \tau+1.
\eeq  
This basis also has a distinguished physical role: it diagonalizes the fusion rules\footnote{This can be seen directly from the Verlinde formula \cite{VERLINDE1988360} for the {\bf rep basis} fusion coefficients
\beq
\mc R_{j_1}\times \mc R_{j_2}=\sum_{j_3}{\mc N_{j_1j_2}}^{j_3}\mc R_{j_3}\qquad\qquad {\mc N_{j_1j_2}}^{j_3}=\sum_{\ell}\frac{{\mc S_{j_1}}^\ell{\mc S_{j_2}}^\ell{\mc S^\dagger_{\ell}}^{j_3}}{{\mc S_0}^\ell}
\eeq
and using the unitarity of $\mc S$.}
\beq
\tmc R_{\tilde j_1}\times\tmc R_{\tilde j_2}=\left({\mc S_0}^{\tilde j_1}\right)^{-1}\delta_{\tilde j_1,\tilde j_2}\tmc R_{\tilde j_1}\qquad\qquad \tmc R_{\tilde j}:=\sum_j{\mc S^\dagger}_{\tilde j}^j\mc R_j
\eeq
Like $\mc T$, $\mc S$ is also a unitary matrix.  
For $SU(2)_k$, $\mc S$ is real and symmetric:
\beq
{\mc S_{j_1}}^{j_2}=\sqrt{\frac{2}{k+2}}\sin\left(\frac{\pi (2j_1+1)(2j_2+1)}{k+2}\right).
\eeq

\subsection{Magic and mana}\label{sect:manarev}
In quantum computation, two important features of a given set of gates are fault-tolerance (the ability to correct for errors within a threshold), and universality (the ability to approximate generic quantum circuits within in a threshold).  These two features do not go hand-in-hand.  For instance, exemplifying the first quality are the {\it Clifford gates} for which there exist many error correcting codes where they can be implemented transversally\footnote{that is they do not propagate errors across differing subsystems between code-blocks. \cite{Preskill:1997uk}}.  We define them shortly below.\\
\\
Let $\{|a\rangle\}_{a=1,\ldots,d}$ be a fixed computation basis of qudits of odd dimension, $d$.  We define generalized Pauli operators (sometimes called the ``shift" and ``clock" operators, respectively) as
\beq
X|a\rangle=|a+1,\,\text{mod}\,d\rangle\qquad\qquad Z|a\rangle=\exp\left(\frac{2\pi i\,a}{d}\right)|a\rangle
\eeq
The Heisenberg-Weyl operators are composed of monomials of $X$ and $Z$:
\beq
{\bf T}_{u,v}:=\exp\left(-\frac{i\pi\,(d+1)uv}{d}\right)Z^{u}X^v\qquad\qquad u,\,v\in\mathbb Z_d
\eeq
Similar operators can be defined for $n$ parties of qudits:
\beq
{\bf T}_{\vec u,\vec v}={\bf T}_{u_1,v_1}\otimes{\bf T}_{u_2,v_2}\otimes\ldots\otimes {\bf T}_{u_n,v_n}\qquad \vec u=\{u_1,u_2,\ldots, u_n\}\in\mathbb Z_d^n\qquad \vec v=\{v_1,v_2,\ldots,v_n\}\in\mathbb Z_d^n
\eeq
Eigenstates of Heisenberg-Weyl operators are called {\it stabilizer states}.  The Clifford group is the set of unitaries that preserve the set of Heisenberg-Weyl operators up to phase
\beq\label{eq:Clifforddef}
\mc C=\{ \mc U \;|\; \mc U^\dagger {\bf T}_{\vec u,\vec v} \,\mc U=e^{i\theta}{\bf T}_{\vec u',\vec v'}\}.
\eeq
While fault-tolerant, $\mc C$ is not universal: a natural consequence of \eqref{eq:Clifforddef} is that the $\mc C$ orbit (possibly along with any projective measurements) of any classical distribution of computational basis states is contained in STAB, the convex hull of stabilizer states.  Moreover, unitary circuits comprised solely of Clifford gates can be efficiently simulated by classic computation (this is the substance of the Gottesmann-Knill theorem \cite{gottesman1998heisenberg,veitch2012negative}).\\
\\
On the other hand, any effort to close this gap will spoil fault-tolerance \cite{eastin2009restrictions}.  For instance, one can add to the set of Clifford gates an additional ``T-gate" (not to be confused with the modular T-matrix, $\mc T$), represented for qutrit systems as
\beq
\mathbb{T}=\text{diag}\left(\exp\left(\frac{-2\pi i}{9}\right),\,1,\,\exp\left(\frac{2\pi i}{9}\right)\right).
\eeq
to achieve a universal quantum circuit.  However, if $\mc C$ can be implemented transversally in a code, then the T-gate cannot.\\
\\
One work-around for this unfortunate situation is to supplement a fault-tolerant, yet non-universal, quantum circuit with a set of {\it resource states} that cannot be prepared with Clifford-gates alone.  One can then, through stabilizer protocols (i.e action of Clifford unitaries, composition with stabilizer states, projective measurements in the computational basis, and partial traces), achieve universal quantum computation.  These states are the {\it magic states} which simply means that they lie outside STAB.    While such states might be prepared noisily, from them one can produce high-quality magic states of arbitrarily high fidelity from stabilizer protocols in what is known as {\it magic distillation} \cite{bravyi2005universal, knill2005quantum, reichardt2006quantum, anwar2012qutrit, campbell2012magic}.
\\
\\
To classify and quantify the ``non-stabilizer-ness" of a state, it is useful to define a {\it magic monotone:} a real map on density matrices that is (i) invariant under $\mc C$, (ii) zero for stabilizer states, and (iii) non-increasing under stabilizer protocols.  For odd and prime dimensional Hilbert spaces\footnote{In this case $\{\vec u,\vec v\}$ belong to the finite vector space $\mathbb Z_{d}^{2n}$.  The same results have been established in odd dimensions with minor modification \cite{gross}, however for simplicity of the discussion we will focus solely on $d$ odd and prime in this paper.}, we can readily use the technology of {\it the discrete Wigner transform} to assign a computable\footnote{We contrast this with another magic monotone, the relative entropy of magic \cite{Veitch:2014aa} defined for a state $\rho$ as
\beq
\theta(\rho)=\min_{\sigma\in\text{STAB}}S(\rho||\sigma)
\eeq
where $S(\rho||\sigma)$ is the relative-entropy.  While this measure matches our criteria, it is nigh-impossible to carry out this minimization practically.} and meaningful measure of magic to pure and mixed states \cite{gross}.  To be more specific, we define the discrete Wigner transform of a state $\rho$ as
\beq
W_{\vec u,\vec v}[\rho]=\frac{1}{d^n}\Tr\left(\rho\,{\bf A}_{\vec u,\vec v}\right)\qquad\qquad {\bf A}_{\vec u,\vec v}=\frac{1}{d^n}{\bf T}_{\vec u,\vec v}\left(\sum_{\vec u',\vec v'}{\bf T}_{\vec u',\vec v'}\right){\bf T}^\dagger_{\vec u,\vec v}
\eeq 
The ${\bf A}_{\vec u,\vec v}$'s (the {\it phase-space point operators}) obey both an orthogonality and a completeness relation
\beq
\frac{1}{d^n}\Tr\left({\bf A}_{\vec u,\vec v}{\bf A}_{\vec u',\vec v'}\right)=\delta_{\vec u,\vec u'}\delta_{\vec v,\vec v'}\qquad\qquad \frac{1}{d^n}\sum_{\vec u,\vec v}{\bf A}_{\vec u,\vec v}=\hat 1
\eeq
and so this decomposition of $\rho$ is well defined and the discrete Wigner transform is normalized:
\beq
\sum_{\vec u,\vec v}W_{\vec u,\vec v}[\rho]=\Tr(\rho)=1
\eeq
Importantly, unlike the density matrix itself, the discrete Wigner transform is only a quasi-probability distribution and so it is allowed to possess negative eigenvalues.  Moreover, there is significance to these negative eigenvalues.  The discrete Hudson's theorem \cite{gross} states that for pure states $W_{\vec u,\vec v}$ is a genuine probability distribution {\it if and only if} that state is stabilizer:
\beq
W_{\vec u,\vec v}[|\psi\rangle\langle \psi|]\geq 0\qquad\Leftrightarrow\qquad |\psi\rangle\text{ is stabilizer}.
\eeq
Thus for pure states we have a sharp diagnosis of a state's magic which is the {\it mana}, defined to be the log negativity of the discrete Wigner transform:
\beq
\mc M(\rho):=\ln\sum_{\vec u,\vec v}\Big| W_{\vec u,\vec v}[\rho]\Big|.
\eeq
For mixed states, this criterion is less-sharp: $\rho\in\text{STAB}$ $\Rightarrow$ $\mc M=0$, but not the other way around.  However, a non-zero $\mc M$ still diagnoses the presence of magic in mixed states.  Physically, $\mc M(\rho)$ lower-bounds the number of resource magic states needed to prepare $\rho$ from a fixed computation basis state and a stabilizer protocol \cite{Gross:2017aa, White:2020zoz}.  Below we list some important facts about $\mc M(\rho)$, which are either obvious or have been proved elsewhere \cite{Veitch:2014aa}:
\begin{itemize}
\item {\bf Positivity:}
\beq
\mc M(\rho)\geq 0
\eeq
\item {\bf Additivity on tensor products:}
\beq
\mc M(\rho_1\otimes \rho_2)=\mc M(\rho_1)+\mc M(\rho_2).
\eeq
\item {\bf Non-increasing on stabilizer protocols:}
\beq
\mc M(\Lambda(\rho))\leq \mc M(\rho)\qquad\qquad \Lambda\text{ is a stabilizer protocol.}
\eeq
\item {\bf Subsystem concavity:} Let $\rho$ be a state on $\mc H=\mc H_A\otimes \mc H_B$ and $\rho_A$ and $\rho_B$ be its reduced states on the respective subsystems. Then
\beq\label{eq:concavity}
\mc M(\rho)\geq\text{max}\left(\mc M(\rho_A),\mc M(\rho_B)\right)\!\!\!\!\qquad\Leftrightarrow\qquad\!\!\!\!\mc M(\rho)\geq x\mc M(\rho_A)+(1-x)\mc M(\rho_B)\;\;\;\;\forall x\in[0,1]
\eeq
which follows directly from the fact that $\mc M$ is non-increasing under partial-traces.
\end{itemize}
In the context of Chern-Simons theory where the dimension of the torus Hilbert spaces of interest are functions of the level, we will label the mana accordingly, i.e. $\mc M_k(\rho)$.  It is important to note that the mana is not a local unitary invariant, and thus a couple of comments are in order.
\begin{itemize}
\item Firstly, the mana associated to a knot or a link is a {\it framing-dependent} object, and thus not strictly topological.  While it is possible to assign a fixed, particular framing of the knots and links in this paper (e.g. a canonical framing) that choice is a somewhat arbitrary resolution.  Instead, the approach we will take in this paper is to define a magic monotone associated to a particular topology by varying over {\it all} framings.  Thus for a state $\rho$ on an $n$-fold tensor product of torus Hilbert spaces, $\otimes_{i=1}^n\mc H_{T^2}$, we define the {\it topological mana} as the minimum mana over the entire framing orbit:
\beq\label{eq:Mtopdef}
{\bf M}_{top}(k;\rho):=\min_{\{m_i\}\in\mathbb Z^n}\mc M_k\left(\left(\otimes_{i=1}^n{\mc T}^{m_i}\right)\cdot\rho\cdot\left(\otimes_{i=1}^n{\mc T^\dagger}^{m_i}\right)\right).
\eeq
The motivation for this definition follows from the interpretation of $\mc M_k(\rho)$ as a lower bound: ${\bf M}_{top}$ bounds the answer to ``how many resource states do I need to prepare $\rho$, in the {\it most optimistic} framing?"  This is somewhat akin to discussions of circuit complexity: while there might be many unitary circuits that construct a given state from a fiducial state, a reasonable definition of complexity should count the gates of the {\it optimal} circuit.  Given only the $PSL(2,\mathbb Z)$ defining relations, finding the minimum in \eqref{eq:Mtopdef} might seem difficult, however, happily, the representation of $\mc T$ acting on the space of integral representations of $\mf g_k$ has finite order.  For instance, for $\mf{su}(2)_k$
\beq
\mc T_{j}=\exp\left(2\pi i\frac{j(j+1)}{k+2}\right).
\eeq
Since $j$ are half-integer, $\mc T$ has order $4k+8$ and thus this minimization can be performed over $\mathbb Z_{4k+8}$ set of framings.

\item Secondly, we need to be clear about the specification of the computational basis with respect to which we define our Clifford group.  
The most natural basis suggested by path-integral preparation is what we have been calling the {\bf rep basis} generated by simple Wilson loop insertions.  Topological mana calculated using this as the computational basis will be denoted as
\beq
{\bf M}_{top}^{(\text{rep})}(k;\rho).
\eeq
As one might guess from the previous section, we will also consider the {\bf Verlinde basis} in this paper.  Although this basis is bit abstract from the perspective of path-integration, it is distinguished by the fusion rules.  We will see in the next section that comparing the mana defined with respect to this computational basis to that in the {\bf rep basis} is a useful way to characterize the magic structure of knots.  The {\bf Verlinde computational basis} will also play a large role for torus links when we discuss long-range magic in section \ref{sect:LRM}.  Topological mana calculated using a {\bf Verlinde computational basis} will be denoted as
\beq
{\bf M}_{top}^{(\text{Ver.})}(k;\rho).
\eeq 
\end{itemize}

\subsection{Abelian Chern-Simons theory}\label{sect:AbelianCS}
For $U(1)_k$ Chern-Simons we can make some very general statements.  This is because for every link state, the wave-functions are completely controlled by the Gauss linking numbers, $\ell_{ij}\,\text{mod}\,k$,
\beq
|\mc L\rangle=\sum_{\{q_i\}=0}^{k-1}\exp\left(\frac{2\pi i}{k}\sum_{i<j}q_iq_j\ell_{ij}\right)|q_1,q_2,\ldots,q_n\rangle
\eeq
This is in a framing such that all self-linking numbers are set to zero.  As emphasized in \cite{Balasubramanian:2018por} (see also \cite{Salton:2016qpp}), such states are a subclass of stabilizer states called {\it weighted graph states.}  Thus it is immediately clear that all link (and knot) states in the Abelian theory possess a trivial topological mana in the {\bf rep basis}:
\beq
{\bf M}_{top}^{(\text{rep})}(k;\rho_{\mc L})={\bf M}_{top}^{(\text{rep})}(k;\rho_{\mc K})=0\qquad\qquad \forall \;\mc L,\;\mc K.
\eeq
In fact this statement is quite a bit stronger.  This is because the modular group acting on the Abelian theory has a Clifford representation\footnote{It's sufficient to show that $\mc S$ and $\mc T$ conjugate $X$ and $Z$ as
\beq
\hat{\mc S}\,X\,\hat{\mc S}^\dagger=Z\qquad \hat{\mc S}\,Z\,\hat{\mc S}^\dagger=X\qquad\hat{\mc T}\,X\,\hat{\mc T}^\dagger=\exp\left(-\frac{i\pi}{k}\right)ZX\qquad\hat{\mc T}\,Z\,\hat{\mc T}^\dagger=Z
\eeq
and so preserve the group of Heisenberg-Weyl operators up to phase.}:
\beq
{\mc S_{q_1}}^{q_2}=\frac{1}{\sqrt k}\exp\left(\frac{2\pi i q_1\,q_2}{k}\right)\qquad\qquad \mc T_{q}=\exp\left(\frac{\pi i\,q^2}{k}\right)
\eeq
(recall that in this theory $\text{dim}\mc H_{T^2}=k$).  Thus the above statement holds also in {\bf Verlinde computation basis}, and {\it along the entire frame orbit}:
\beq
\mc M_k^{(\text{rep/Ver})}(\rho_{\mc L})=\mc M_k^{(\text{rep/Ver})}(\rho_{\mc K})=0\qquad\qquad \forall\,\mc L,\,\mc K,\;\;\forall\,\text{framings}.
\eeq
Since the mana is non-increasing under partial traces, these statements will continue to hold when we discuss ``long-range mana" in section \ref{sect:LRM}.  Thus we see that the magic captures more than topology; it also requires some non-Abelian structure\footnote{There are also some particular non-Abelian Chern-Simons theories whose path integrals produce stabilizer states \cite{schnitzer2019clifford} however these are quite special and while they are characterized by a non-Abelian group, e.g. $SU(N)_1$, their anyon statistics are Abelian.}.  This also emphasizes a point we hinted at in the introduction: the magic is a characterization of a state that is quite distinct from entanglement.  The states prepared by Abelian Chern-Simons theory generically have both short-range and long-range entanglement, however we see that from this diagnostic they are ``simple."\\
\\
For the rest of this paper we will be focused on states in the simplest non-Abelian theory: $SU(2)_k$ Chern-Simons theory.  The Hilbert space of a single torus is $k+1$ dimensional so given the discussion in section \ref{sect:manarev}, the low-lying values of $k$ of relevance here are $k=2,4,6,10,$ and $12$.  The $k=2$ and $k=4$ theories prepare qutrits and ququints (respectively), which are well-trodden domains in the magic and mana literature.  This is where we do the lion's share of computations in the paper, however for a handful of specific knots we will perform computations ranging through the allowed values of $k$ up to $k=12$ and for a handful of links, up to $k=6$.

\section{Knots}\label{sect:knots}

\begin{figure}[h!]
\centering
\begin{subfigure}[b]{.7\textwidth}
\includegraphics[width=10cm]{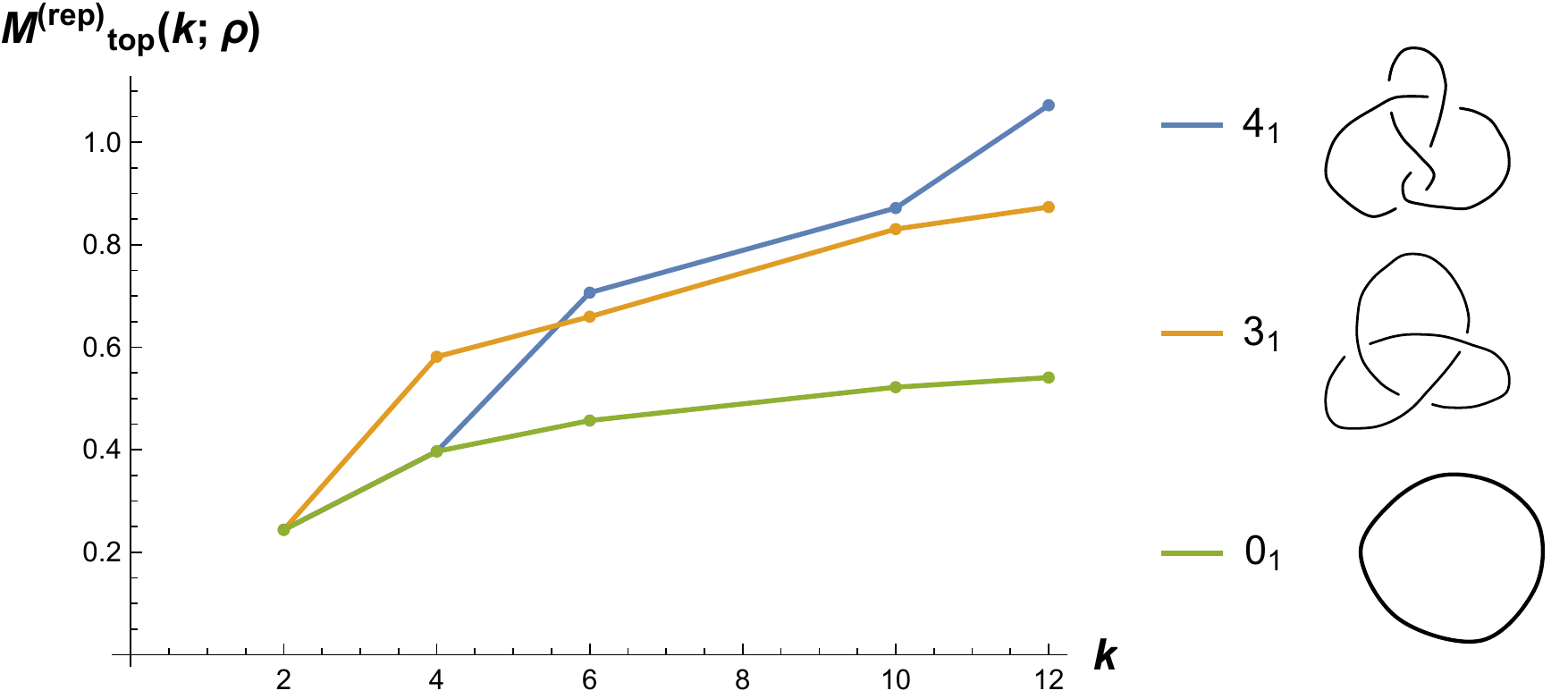}
\caption{\small{\textsf{Plots of ${\bf M}_{top}$ in the {\bf rep computational basis}.  While they share the same non-zero value at $k=2$, they diverge as $k$ (and $\text{dim}\mc H_{T^2}$) increases.}}}
\end{subfigure}

\begin{subfigure}[b]{.7\textwidth}
\includegraphics[width=10cm]{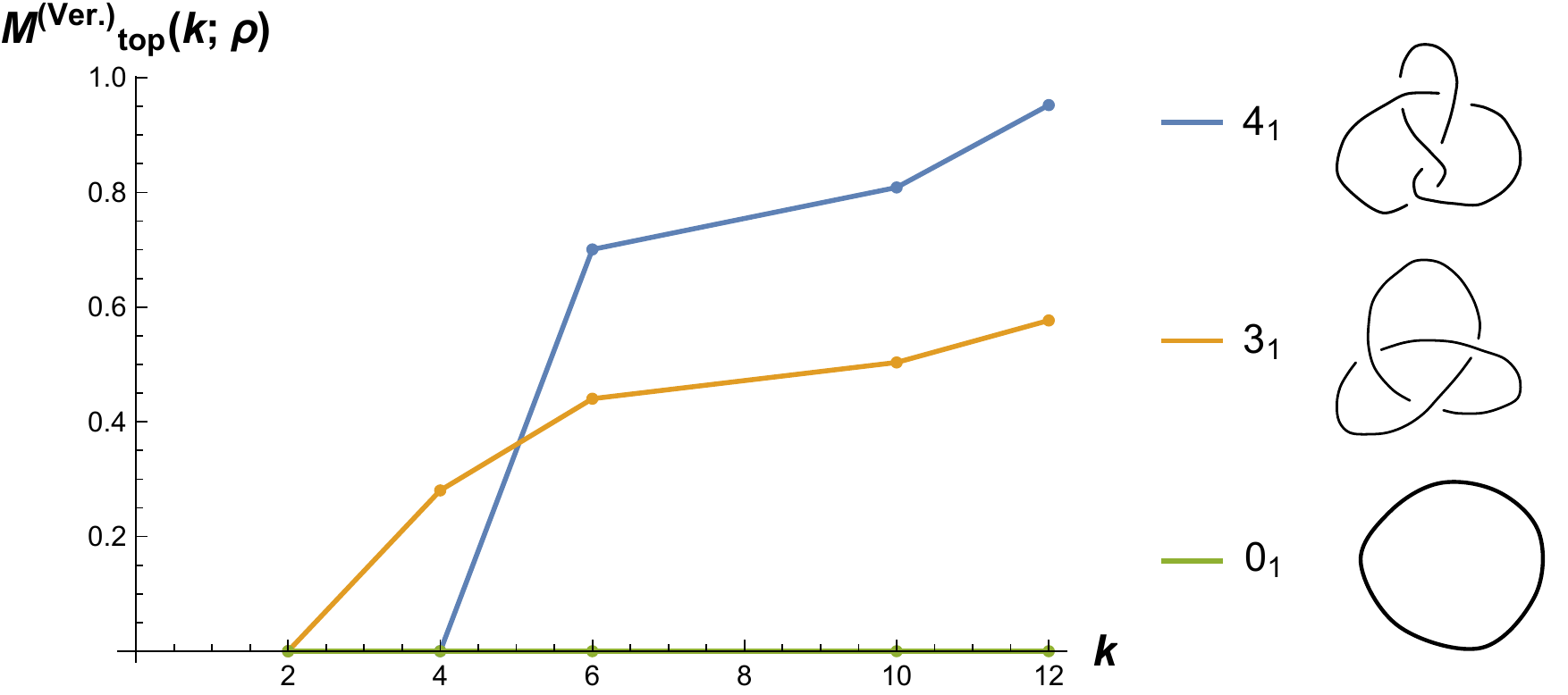}
\caption{\small{\textsf{Plots of ${\bf M}_{top}$ in the {\bf Verlinde computational basis}.  While they are all zero at $k=2$, only the unknot remains zero for all allowed $k$.}}}
\end{subfigure}
\caption{\small{\textsf{The dependence of ${\bf M}_{top}$ over the allowed values of $k$ for a few distinguished knots: the unknot ($0_1$), the trefoil ($3_1$), and the figure-eight knot ($4_1$).  The contours are interpolating lines only meant to guide the eye, ${\bf M}_{top}$ is only defined for $k=$2, 4, 6, 10, and 12 in this plot.}}}
\label{fig:knotmanaplot}
\end{figure}

We begin with a description of the topological mana for a large list of knots.  The central tool for constructing the wave-functions for these states is the calculation of the knot invariants from conformal block monodromies of the chiral $SU(2)_k$ WZW conformal field theory.  We use the procedure\footnote{Or more accurately a variant mentioned in \cite{Witten:1988hf} and further developed in \cite{Balasubramanian:2018por}} of \cite{Kaul:1993hb,Kaul:1998ye,Kaul:1991np}, by evaluating the knot invariant as the trace of braid-operators on the punctured two-sphere Hilbert space, $\mc H_{S^2}$.  This method works equally well for links.  For purposes of being self-contained we provide a lightning review of the method in appendix \ref{app:braidcomp}.
\\
\\
For the first result, in the qutrit ($k=2$) theory, of the 64 knots we considered {\it every knot} had non-zero mana in the {\bf rep basis}.  Moreover, this mana was {\it identical} amongst every knot:
\beq
{\bf M}^{(\text{rep.})}_{top}(2;\rho_{\mc K})=\ln\left(\frac{1}{3}\left(1+2\sqrt 2)\right)\right)\approx.243842\qquad\qquad \forall\;\mc K\in\text{list}_{\text{knots}}
\eeq 
where $\text{list}_{\text{knots}}$ is enumerated in appendix \ref{app:listknots}.  This list includes a few well-known knots, e.g. the unknot ($0_1$ in Rolfsen notation), the trefoil ($3_1$), and the figure eight ($4_1$).  
This implies that every knot state in the {\bf rep basis} is non-stabilizer.  Thus there is something inherently magical about the $S^3$ path-integral preparation of states, even for a single unknotted boundary.  
For the unknot the origin of this non-Clifford preparation is clear: the $0_1$ link invariant can be obtained by gluing an empty solid torus to the link complement with an ${\mc S}$ twist (to fill in the $S^3$) and so operationally one can prepare $|0_1\rangle$ by applying the modular S operator, $\hat{\mc S}$, to the computational $|0\rangle$ state:
\beq
\langle j|0_1\rangle={\mc S_0}^j\qquad\Rightarrow\qquad |0_1\rangle=\hat{\mc S}|0\rangle
\eeq
So $\hat{\mc S}$ is precisely the non-Clifford unitary responsible for the non-zero mana.  This can be made more explicit in the {\bf Verlinde basis} where we can verify that the topological mana is zero for all valid $k$'s:
\beq
{\bf M}^{(\text{Ver.})}_{top}(k;\rho_{0_1})=0
\eeq 
Astoundingly, we find that for $k=2$, this is also true of {\it all knots} in $\text{list}_{\text{knots}}$:
\beq
{\bf M}^{(\text{Ver.})}_{top}(2;\rho_{\mc K})=0\qquad\qquad \forall\;\mc K\in\text{list}_{\text{knots}}
\eeq
Since these are pure states, there are a couple of interesting implications (citing the 64 knots in $\text{list}_{\text{knots}}$ as evidence) one can draw from this:
\begin{itemize}
\item All $SU(2)_2$ knots states possess a framing such that they are stabilizer in the {\bf Verlinde basis}.  Equivalently, knot states are non-stabilizer in the {\bf rep basis} only up to framing and a single application of $\hat{\mc S}$.\\
\item Every $SU(2)_2$ knot state can be prepared by a combination of framing and a stabilizer protocol acting on the unknot.
\end{itemize}  
These two statements seem to be special for $SU(2)_2$: while knots still generically have non-zero ${\bf M}^{(\text{rep})}_{top}$ for $k=4$, this value varies from knot to knot.  In the {\bf Verlinde basis}, ${\bf M}_{top}^{(\text{Ver.})}(4;\rho_{\mc K})=0$ for many knots, however there are also many examples of non-zero mana, e.g. the trefoil
\beq
{\bf M}_{top}^{(\text{Ver.})}(4;\rho_{3_1})\approx 0.28036
\eeq
A table of the $k=4$ topological mana's for knots is provided in appendix \ref{app:listknots}.  For $k>4$ we plot the dependence of ${\bf M}_{top}$ on $k$ in both the {\bf rep} and {\bf Verlinde bases} for a handful of knot states in figure \ref{fig:knotmanaplot}.\\

\section{Long-range magic}\label{sect:LRM}
\begin{figure}[h!]
\centering
\begin{subfigure}[c]{.1\textwidth}
\includegraphics[width=1cm]{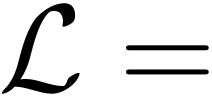}
\end{subfigure}
\begin{subfigure}[c]{.125\textwidth}
\includegraphics[width=1.5cm]{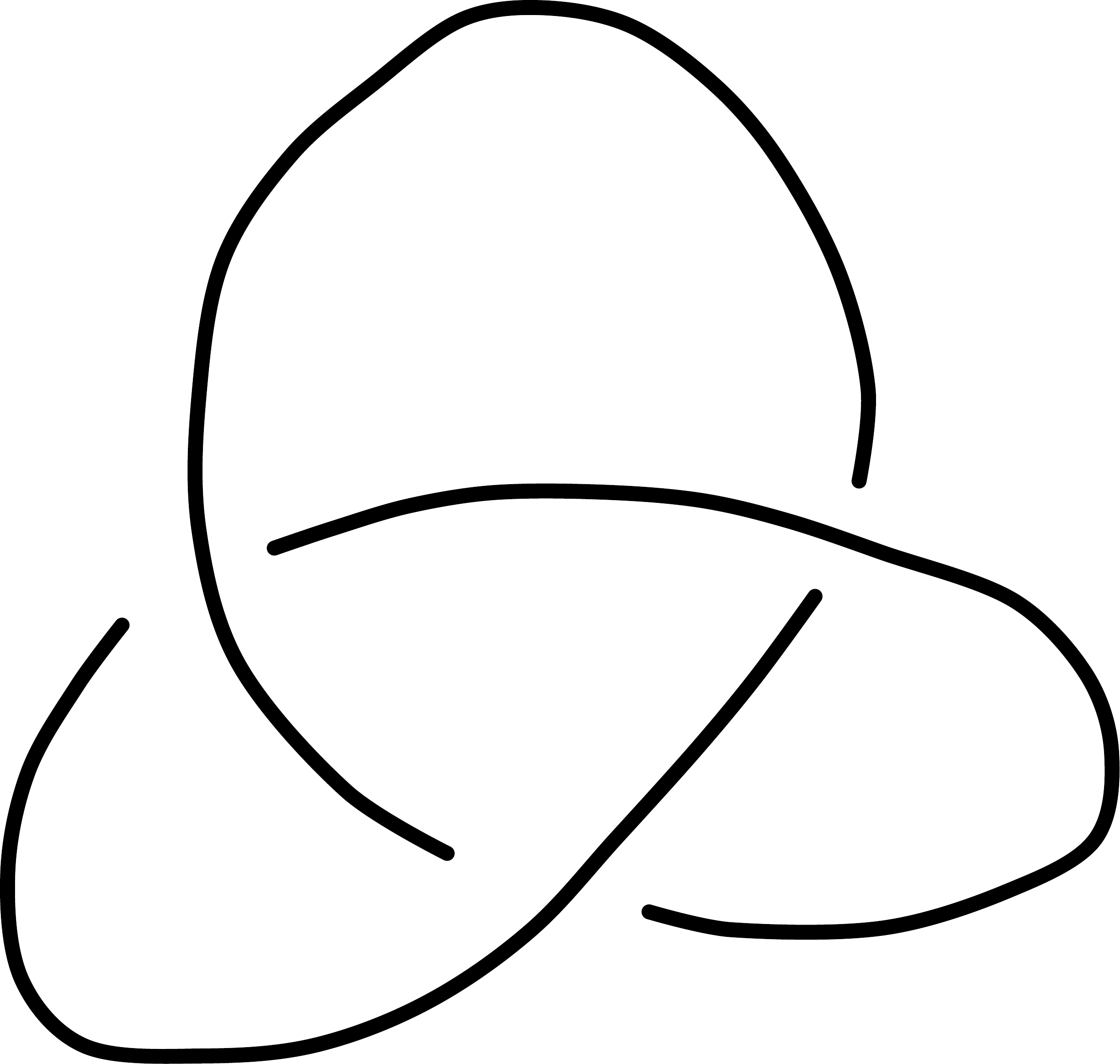}
\end{subfigure}
\begin{subfigure}[c]{.125\textwidth}
\includegraphics[width=1.5cm]{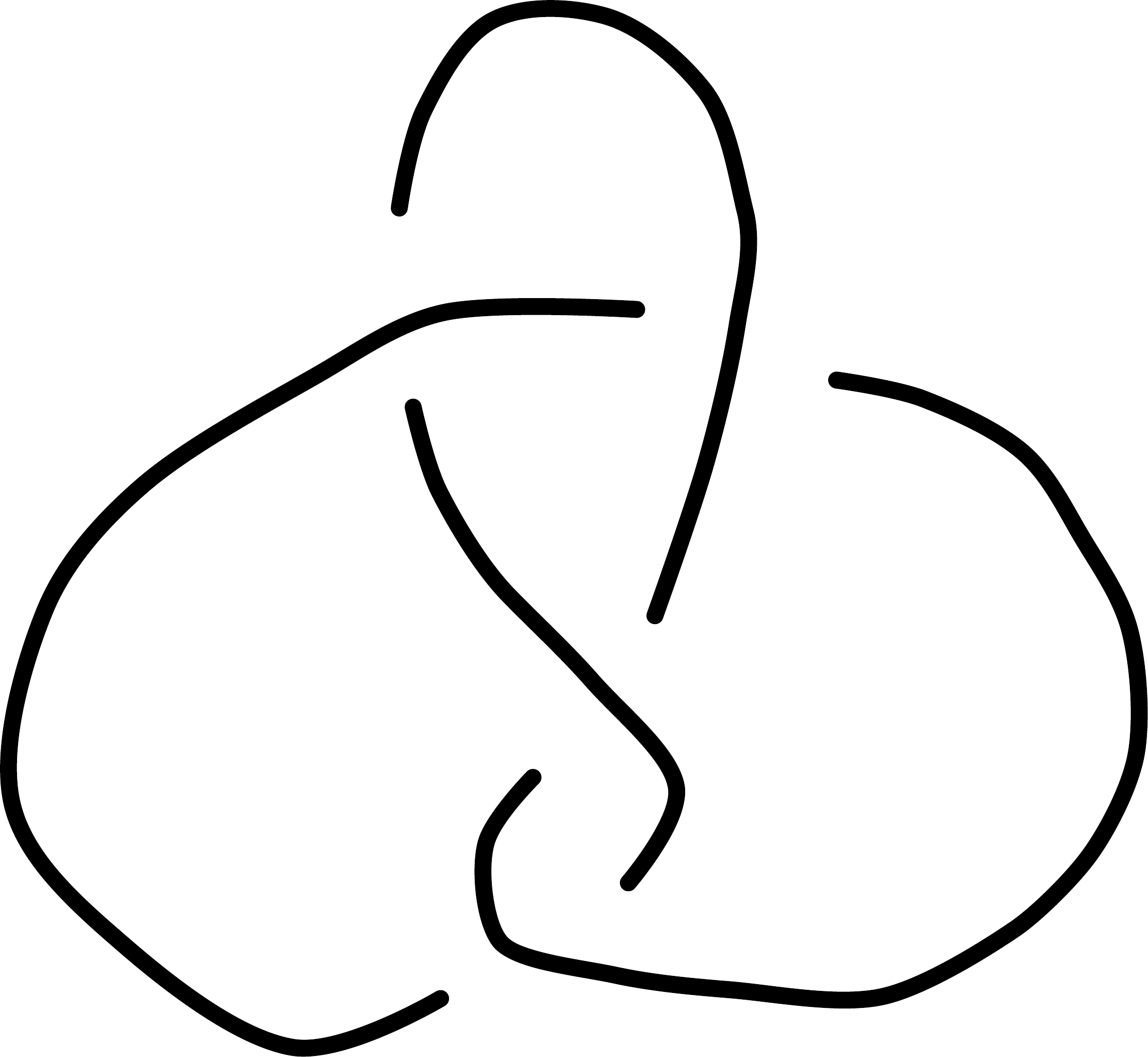}
\end{subfigure}
\caption{\small{\textsf{When $\mc L=\mc K_1\sqcup\mc K_2$ with $\mc K_1$ (here the trefoil $3_1$) and $\mc K_2$ (here the figure eight $4_1$) disconnected the wave-function factors: $\Psi_{\mc L}=\Psi_{\mc K_1}\Psi_{\mc K_2}$ and so the state is a product state.}}}
\label{fig:disconnlink}
\end{figure}

We now move onward to our secondary focus of this paper which is how magic can be distributed across subsystems.  Since the subsystems of this paper are associated to distinct boundaries with no associated length-scale, we will be discussing {\it long-range magic.}  For a given bipartition of the torus Hilbert spaces, $\mc H=\mc H_A\otimes \mc H_{B}$, we measure this by comparing the mana of the global state and mana of the state reduced upon its subfactors\footnote{Note that while the spectrum of the $\rho_A$ is the same as $\rho_{B}$, unlike entanglement entropy, $\mc M$ is not necessarily complementary since the local unitaries implementing the Schmidt decomposition can be of independent complexity.}, in what we name the {\it long-range mana}:
\beq
L_k(\rho_{\mc L})=\mc M_k(\rho_{\mc L})-\mc M_k(\rho_{A})-\mc M_k(\rho_{B})\qquad \rho_A=\Tr_{\mc H_B}|\mc L\rangle\langle \mc L|\qquad \rho_B=\Tr_{\mc H_A}|\mc L\rangle\langle \mc L|
\eeq
$L_k$ has appeared previously as the ``connected component of mana" in \cite{White:2020zoz} or the ``global magic"\footnote{We stress to the reader that in this paper we refer to the magic of a pure state on the full bipartite Hilbert space also as the {\it global magic}.  We hope that this does not cause confusion.} in \cite{Sarkar_2020}.  It provides a coarse measure of a state's magic stored in the correlations between subsystems.  Indeed, since the mana is additive for factorized states, $\mc M_k(\rho_A\otimes \rho_{B})=\mc M_k(\rho_A)+\mc M_k(\rho_{B})$, uncorrelated subsystems share no long-range mana, as expected.  While it is clear that the motivation of $L_k$ is to ``subtract off" the magic of local (to $\mc H_A$ or $\mc H_B$) unitaries preparing $\rho_{\mc L}$, it is important to note that {\it $L_k$ is not local unitary invariant.}  Importantly, $L_k$ might vary with framing.  Given this, we then define the {\it topological long-range mana} as the minimum of $L_k$ over the frame orbit:
\beq
{\bf L}_{top}(k;\mc L):=\min_{\{m_i\}\in\mathbb Z^n}L_k\left(\left(\otimes_{i=1}^n{\mc T}^{m_i}\right)\cdot\rho_{\mc L}\cdot\left(\otimes_{i=1}^n{\mc T^\dagger}^{m_i}\right)\right)
\eeq
Note that links composed of disconnected components (for example figure \ref{fig:disconnlink}) are product states regardless of framing and thus have no long-range magic between them:
\beq
{\bf L}_{top}(k;\mc L=\mc K_1\sqcup\mc K_2)=0
\eeq
\\
We note that since there always exists a framing such that $L_k\leq{\bf M}_{top}$ (e.g. the framing that realizes $\mc M_k={\bf M}_{top}$), the topological long-range mana is always bounded above by the topological mana of the global pure state:
\beq
{\bf L}_{top}\leq {\bf M}_{top}
\eeq
We will be interested in comparing ${\bf L}_{top}$ with ${\bf M}_{top}$, however since both involve minimizations over the frame orbit, we have to be careful.  In particular, the framing minimizing $\mc M_k$ could be different than the framing minimizing $L_k$.  This could happen when an increase in $\mc M_k(\rho_A)+\mc M_k(\rho_{B})$ is under-compensated by an increase in $\mc M_k(\rho_{\mc L})$.  However, importantly, when ${\bf L}_{top}$ matches the topological mana, a direct comparison is possible:\\\\
{\bf Claim:} {\it if ${\bf L}_{top}={\bf M}_{top}$ then any framing minimizing $\mc M_k(\rho_{\mc L})$ also minimizes $L_k(\rho_{\mc L})$}.\\
\\
{\bf Proof:} Let the framing minimizing $L_k$ be called $f_L$.  If there is more than one, choose one.  Let us also choose a framing minimizing $\mc M_k$ and denote it $f_{\mc M}$.  If $f_L=f_{\mc M}$ then we are done, so let us suppose they are different.  Then by definition of $f_L$ minimizing $L_k$:
\beq
0\leq L_k\left(\rho_{\mc L}^{(f_{\mc M})}\right)-L_k\left(\rho_{\mc L}^{(f_L)}\right)=\mc M_k\left(\rho_{\mc L}^{(f_{\mc M})}\right)-\mc M_k\left(\rho_A^{(f_{\mc M})}\right)-\mc M_k\left(\rho_{B}^{(f_{\mc M})}\right)-L_k\left(\rho_{\mc L}^{(f_L)}\right)
\eeq
However by supposition $\mc M_k(\rho_{\mc L}^{(f_{\mc M})})={\bf M}_{top}(\rho_{\mc L})={\bf L}_{top}(\rho_{\mc L})=L_k(\rho_{\mc L}^{(f_L)})$.  We then have
\beq
\mc M_k\left(\rho_A^{(f_{\mc M})}\right)+\mc M_k\left(\rho_{B}^{(f_{\mc M})}\right)\leq 0
\eeq
However since the mana is non-negative, the only way this can be possible is for $\mc M_k\left(\rho_A^{(f_{\mc M})}\right)=\mc M_k\left(\rho_{A^c}^{(f_{\mc M})}\right)=0$.  Thus
\beq
L_k\left(\rho_{\mc L}^{(f_{\mc M})}\right)=\mc M_k\left(\rho_{\mc L}^{(f_{\mc M})}\right)={\bf M}_{top}(k;\rho_{\mc L})={\bf L}_{top}(k;\rho_{\mc L})
\eeq
and so achieves the minimum.\\\\
As an example to keep in mind, let us look at the topological long-range magic of the Hopf link, L2a1.  The state in the {\bf rep basis} (including arbitrary framing) is
\beq
|\text{L2a1}\rangle=\frac{1}{\sqrt{k+1}}\sum_{j_1,j_2}{\mc S_{j_1}}^{j_2}{\mc T_{j_1}}^{m_1}{\mc T_{j_2}}^{m_2}|j_1,j_2\rangle
\eeq
As a pure state, $|\text{L2a1}\rangle$ is magical in both computational bases. E.g. for $k=2$
\beq
{\bf M}_{top}^{(\text{rep})}(2;\rho_{\text{L2a1}})\approx0.69098\qquad\qquad {\bf M}_{top}^{(\text{Ver})}(2;\rho_{\text{L2a1}})\approx 0.398092.
\eeq
However, reducing this state down to a either component leaves it maximally mixed 
\beq
\rho_{\text{L2a1},red}=\frac{1}{k+1}\sum_{j}|j\rangle\langle j|
\eeq
and thus has no magic in any basis, any framing, and any $k$.  Thus all of the mana is long-range:
\beq
{\bf L}^{(\text{rep/Ver})}_{top}(k;\rho_\text{L2a1})={\bf M}_{top}^{(\text{rep/Ver})}(k;\rho_{\text{L2a1}}).
\eeq

\subsection{Torus links}

The above example of the Hopf link extends nicely to an entire class of links called {\it torus links} (of which the Hopf link is the simplest example).  We recall that a torus link is a link whose strands can all be embedded in a non-intersecting manner on the surface of a single $T^2$.  Because of this fact, their wave-functions (up to framing and normalization, $\mc N$) can be evaluated via the fusion of Wilson loop operators acting on the torus \cite{Isidro:1992fz, Labastida:2000yw, Brini:2011wi}.\\
\\
A torus link is labelled by two integers $(P,Q)$ with $\text{gcd}(P,Q)=n$ the number of components of the link.  The fusion of the Wilson loop operators leads to wave-functions in the {\bf rep basis} of the form
\beq\label{eq:LpqWF}
|\mc L(P,Q)\rangle=\mc N^{-1/2}\sum_{\ell_1,\ell_2}\sum_{j_1,\ldots,j_n}\frac{1}{\left({\mc S_0}^{\ell_1}\right)^{n-1}}{\mc S_{j_1}}^{\ell_1}{\mc S_{j_2}}^{\ell_1}\ldots{\mc S_{j_n}}^{\ell_1}{\mc S_{\ell_1}}^{\ell_2}J_{\ell_2}(P/n,Q/n)\,|j_1,j_2,\ldots,j_n\rangle
\eeq
where $J_{\ell}(P/n,Q/n)$ is the colored Jones polynomial of the $(P/n,Q/n)$ torus {\it knot} resulting from the fusion.  Note that the framing of \eqref{eq:LpqWF} generally differs from that arising from the procedure outlined in Appendix \ref{app:braidcomp}.  Fixing the above framing, these states are most simply expressed in the {\bf Verlinde basis} where the fusion rules are diagonalized:
\beq
|\mc L(P,Q)\rangle=\mc N^{-1/2}\sum_{\ell_1,\ell_2}\frac{1}{\left({\mc S_0}^{\ell_1}\right)^{n-1}}{\mc S_{\ell_1}}^{\ell_2}J_{\ell_2}(P/n,Q/n)|\tilde\ell_1,\tilde\ell_1,\ldots,\tilde\ell_1\rangle
\eeq
As an immediate consequence, in this basis and framing, the reduced density matrix of $|\mc L(P,Q)\rangle$ reduced on any subset of components is completely separable:
\beq
\rho_{\mc L(P,Q),red}=\sum_{\ell}\left(\frac{1}{\left({\mc S_0}^{\ell}\right)^{2n-2}}\sum_{\ell_1,\ell_2}{\mc S_{\ell}}^{\ell_1}{\mc S^\ast_{\ell}}^{\ell_2}J_{\ell_1}(P/n,Q/n)J^\ast_{\ell_2}(P/n,Q/n)\right)|\tilde\ell,\ldots,\tilde\ell\rangle\langle\tilde\ell,\ldots,\tilde\ell|
\eeq
(we're allowed to be agnostic about which subset is traced out since the expression is the same for {\it all} non-trivial, proper subsets).  This reduced density matrix is a classical mixture of computational basis states and so obviously lies in STAB and its mana trivially vanishes.  Thus torus link states prepared in this particular basis and this particular framing provide a concrete example of long-range magic:
\beq
L_k(\rho_{\mc L(P,Q)})=\mc M_k(\rho_{\mc L(P,Q)})
\eeq
This magic is both {\it intrinsically non-local} and {\it fragile}, collapsing upon tracing out of any proper subfactors.  Although this doesn't necessarily guarantee\footnote{For instance, \eqref{eq:topMtopLRMtoruslinks} is not guaranteed simply by the GHZ-like entanglement structure of torus link states.  In particular, states of the form
\beq\label{eq:randomtoruslink}
|\psi\rangle=\sum_{\ell}\sum_{j_1,\ldots, j_n}{(\mc U^{(m_1)})_{j_1}}^{\ell}\ldots{(\mc U^{(m_n)})_{j_n}}^{\ell}\lambda_\ell|j_1,\ldots,j_n\rangle\qquad\qquad \{\lambda_\ell\in\mathbb C\;\big|\;\sum_\ell |\lambda_\ell|^2=1\}
\eeq
(where $\hat{\mc U}^{(m)}=\hat{\mc T}^{m}\hat{\mc S}$ for the {\bf rep basis}, including framing, and $\hat{\mc U}^{(m)}=\hat{\mc S}\hat{\mc T}^m\hat{\mc S}$ for the {\bf Verlinde basis}, including framing) typically have ${\bf L}_{top}\neq{\bf M}_{top}$ for generic coefficents $\{\lambda_\ell\}$ so \eqref{eq:topMtopLRMtoruslinks} is likely special to torus links.  Regardless, such states as \eqref{eq:randomtoruslink} also have entirely long-range magic up to local unitary.  Since our motivation for defining ${\bf L}_{top}$ is to {\it roughly} gauge the magic not captured by local unitaries, whether or not \eqref{eq:topMtopLRMtoruslinks} is satisfied is somewhat putting the cart before the horse.}
that the topological long-range mana saturates its upper bound, ${\bf L}_{top}={\bf M}_{top}$ our $k=2$ and $k=4$ numerics for the torus links in tables \ref{table:2linksk2} and \ref{table:2linksk4} in appendix \ref{app:linktable} suggest that this is true, interestingly, in {\it both} computational bases:
\beq\label{eq:topMtopLRMtoruslinks}
{\bf L}_{top}^{(\text{Ver})}(k;\rho_{\mc L(P,Q)})={\bf M}_{top}^{(\text{Ver})}(k;\rho_{\mc L(P,Q)})\qquad{\bf L}_{top}^{(\text{rep})}(k;\rho_{\mc L(P,Q)})={\bf M}_{top}^{(\text{rep})}(k;\rho_{\mc L(P,Q)})
\eeq
While torus links are useful examples, more generally we will only be able to probe a link's long-range magic through computing ${\bf L}_{top}$ numerically.  This only provides a course measure of a link's long-range magic.  This is because $\rho_A$ and $\rho_{B}$ are mixed and therefore ${\mc M}_k(\rho_{A})$ and $\mc M_k(\rho_B)$ only lower bound their magic.  Thus for generic links, it is sometimes helpful to rephrase this as ${\bf L}_{top}(k;\rho_{\mc L})<{\bf M}_{top}(k;\rho_{\mc L})$ indicating the presence of {\it residual subsystem magic} which is an obstruction to simplifying $\rho_{\mc L}$ by reducing on one of its components.
\\
\\
To make this investigation concrete, in what follows we consider a set of two links, $\text{list}_{\text{two-links}}$, 
that are enumerated in appendix \ref{app:linktable} and compute the topological long-range mana between their components.  As mentioned above, this a course measure of the states long-range magic but it is enough to establish some key results.  

\subsection{Two links}\label{sect:2links}
\begin{figure}[h!]
\centering
\begin{subfigure}[b]{\textwidth}
\centering
\includegraphics[width=15cm]{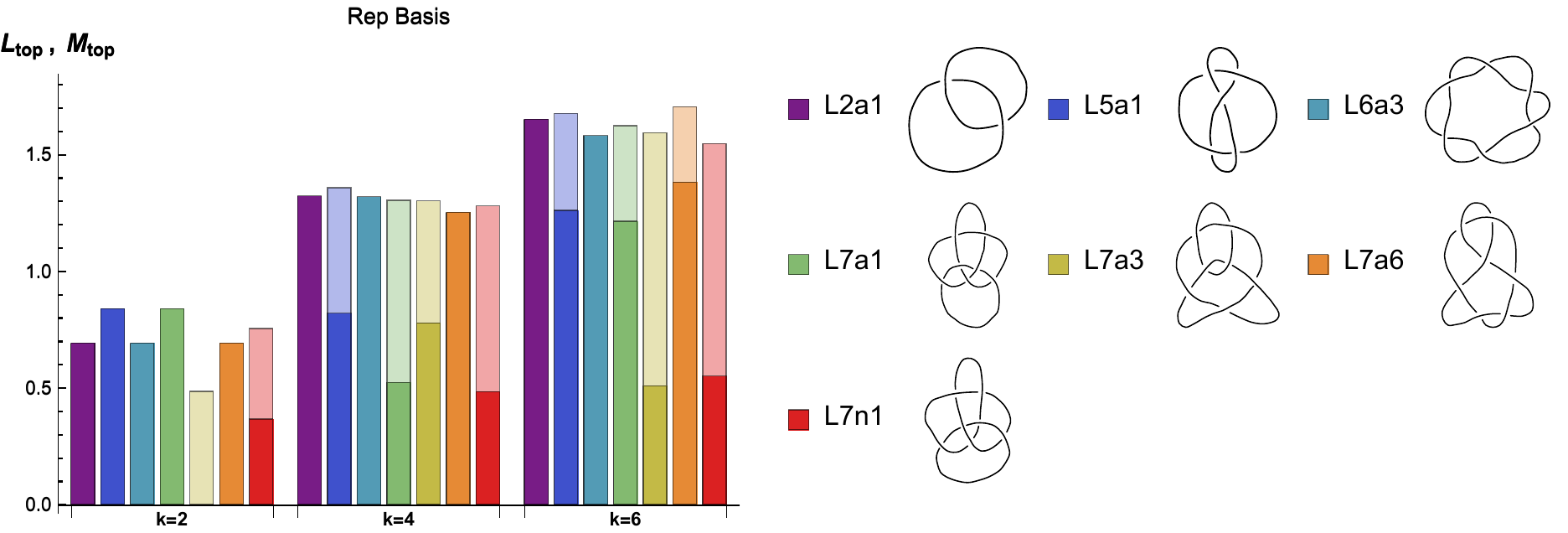}
\caption{\small{\textsf{}}}
\end{subfigure}

\begin{subfigure}[b]{\textwidth}
\centering
\includegraphics[width=15cm]{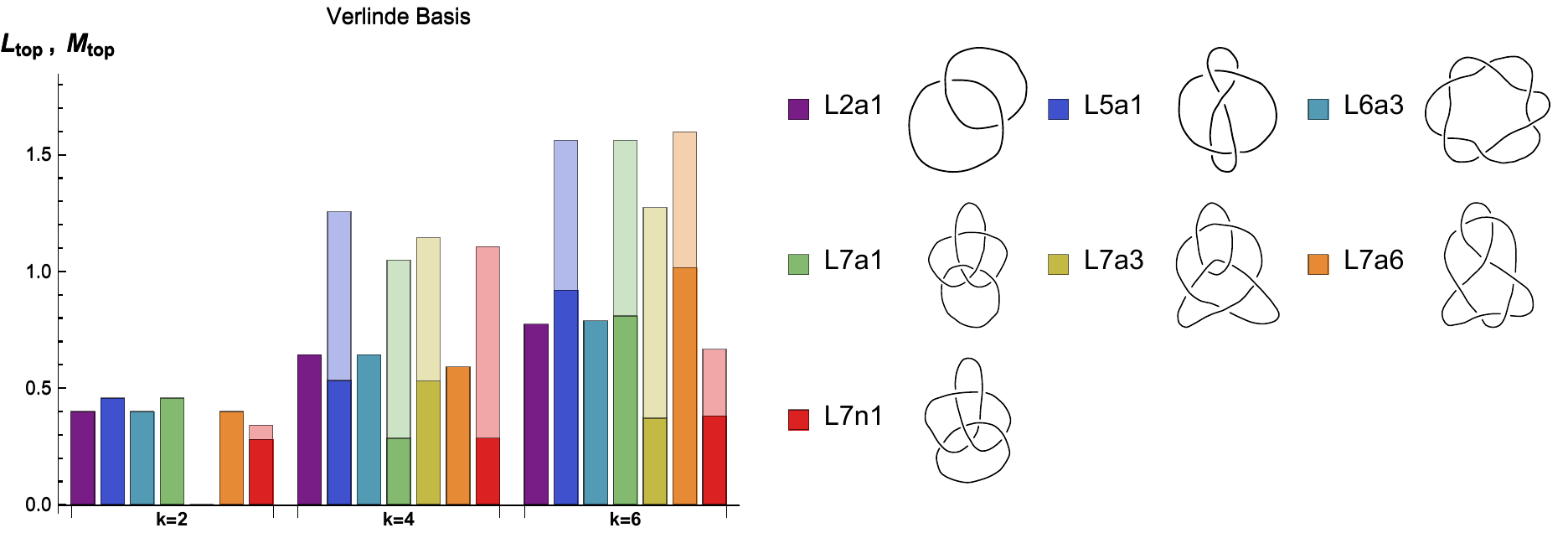}
\caption{\small{\textsf{}}}
\end{subfigure}
\caption{\small{\textsf{The topological long-range mana, ${\bf L}_{top}$, is charted for seven different two-links.  Behind each bar, that link state's full topological mana, ${\bf M}_{top}$, is plotted at half-opacity.  This done for $k=2,4,$ and $6$ (i.e. $\text{dim}\mc H=3,5,$ and $7$, respectively) in both the {\bf rep basis} (subfigure (a)) and the {\bf Verlinde basis} (subfigure (b)).  Note that only for the two torus links, L2a1 and L6a3, are ${\bf L}_{top}$ and ${\bf M}_{top}$ always equal.}}}
\label{fig:2linkLRMbar}
\end{figure}

\begin{figure}[h!]
\centering
\begin{subfigure}[t]{.45\textwidth}
\centering
\includegraphics[width=8cm]{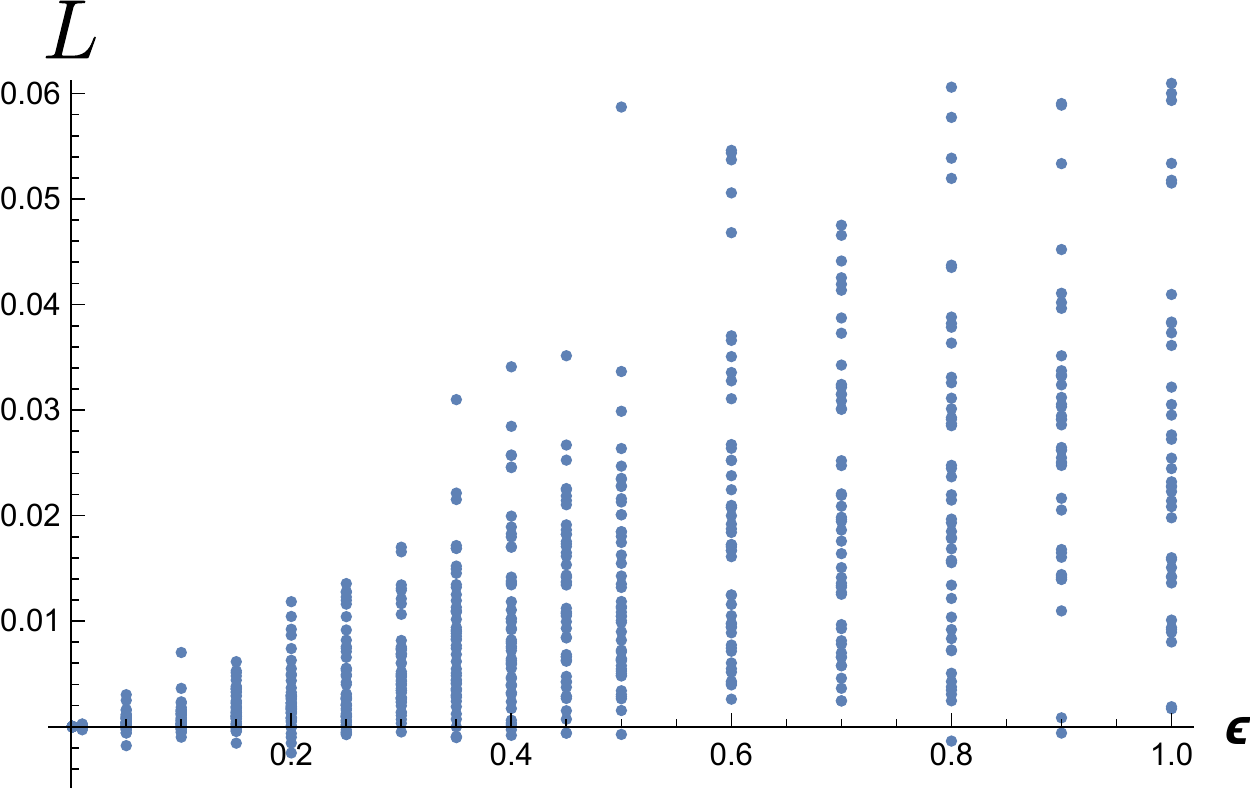}
\caption{\small{\textsf{The plot of $L=\mc M(\rho)-\mc M(\rho_A)-\mc M(\rho_B)$ for $\epsilon\in\{0.001,0.01,0.05,0.1,0.15,0.2,0.25,0.3,0.35,$
$0.4,0.45,0.5,0.6,0.7,0.8,0.9,1\}$.}}}
\end{subfigure}
\qquad
\begin{subfigure}[t]{.45\textwidth}
\centering
\includegraphics[width=8cm]{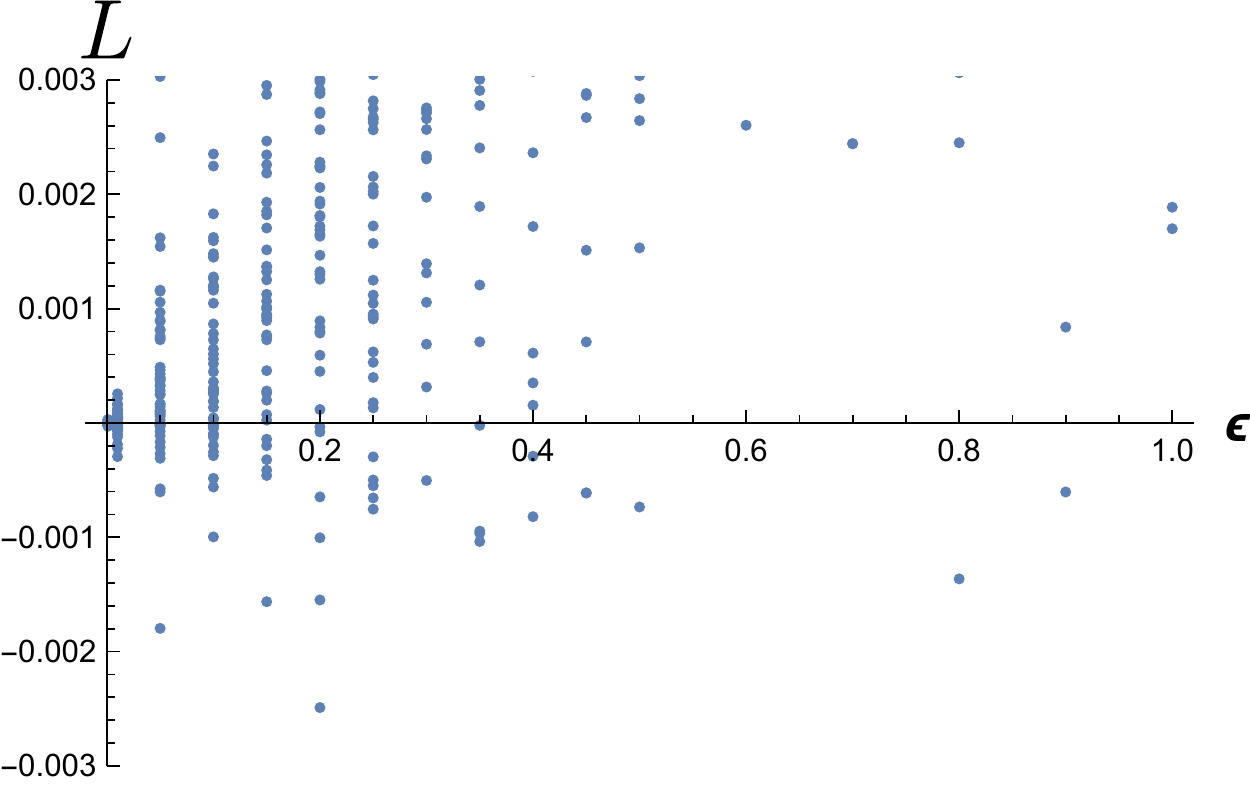}
\caption{\small{\textsf{The same plot with $L$ cutoff between $-0.003$ and $0.003$ to emphasize the negative values. }}}
\end{subfigure}
\caption{\small{\textsf{The long range mana for a collection of pseudo-random ``close-to-product" states of two qutrits: $\mc H_A\otimes\mc H_B=\mc H_3\otimes \mc H_3$.  The states in question are pure and of the form  $|\psi\rangle=|\psi^{(A)}_{p.r.}\rangle_A\otimes|\psi^{(B)}_{p.r.}\rangle_B+\epsilon |\delta\psi_{p.r.}\rangle_{AB}$.  with coefficients in the computational basis given by complex numbers generated pseudo-randomly by Mathematica (i.e. 3 each for $|\psi_{p.r.}^{(A)}\rangle_A$ and $|\psi_{p.r.}^{(B)}\rangle_B$ and 9 for $|\delta\psi_{p.r.}\rangle$).  The state is then normalized.  50 such states are generated for a given $\epsilon$ and $L=\mc M(\rho)-\mc M(\rho_A)-\mc M(\rho_B)$ is computed.  Note that as $\epsilon$ increases, the density of negative $L$'s decreases.  When $\epsilon\sim1$ the typical $|\psi\rangle$ is not close to a product state.}}}
\label{fig:closetoprodplot}
\end{figure}

Starting in the qutrit $k=2$ theory, of the two-links that we considered, all but a handful have a non-zero ${\bf L}^{(\text{rep})}_{top}$ in the {\bf rep basis}.  For most of these links this topological long-range mana is equal to the link's global topological mana ${\bf L}_{top}^{(\text{rep})}={\bf M}_{top}^{(\text{rep})}$ which suggests that the magic of link states is generically long-range.\\
\\
Let us also briefly examine the handful of two-links with vanishing ${\bf L}_{top}^{(\text{rep})}$.  In table \ref{table:2linksk2} of appendix \ref{app:linktable}, these are L7a3, L9a9, L9a38, L9n5, L9n18, and L9n19.  For all six of these link states, the framing minimizing the global mana also minimizes each subsystem mana.  Furthermore, the global magic is completely encapsulated in the subsystem mana:  
\beq
{\bf M}_{top}^{(\text{rep})}(2;\rho_{\text{L7a3}})={\bf M}_{top}^{(\text{rep})}(2;\rho_{\text{L7a3},A})+{\bf M}_{top}^{(\text{rep})}(2;\rho_{\text{L7a3},B})
\eeq
The values for the residual subsystem manas are equal, ${\bf M}_{top}^{(\text{rep})}(2;\rho_{\text{L7a3},A})={\bf M}_{top}^{(\text{rep})}(2;\rho_{\text{L7a3},B})\approx\ln\left(\frac{1}{3}(1+2\sqrt 2)\right)$ and familiar from the topological mana of qutrit knot states: it appears that the residual subsystem mana is captured by a single $\hat{\mc S}$ operation.  This interpretation is corroborated by viewing these links in the {\bf Verlinde basis} where, indeed, both the global topological mana and the residual sub-system mana vanish:
\beq
{\bf M}_{top}^{(\text{Ver})}(2;\rho_{\text{L7a3}})=0\qquad\qquad{\bf M}_{top}^{(\text{Ver})}(2;\rho_{\text{L7a3},A})={\bf M}_{top}^{(\text{Ver})}(2;\rho_{\text{L7a3},B})=0
\eeq
This is true for all six of links mentioned above.  Although these states are not product states, from the perspective of magic, these handful of qutrit link states appear to be no more complex than the tensor product of simple knots.\\
\\
For the ququint $k=4$ theory, the structure of long-range magic appears to be more varied.  While a few links have ${\bf L}_{top}={\bf M}_{top}$, most links possess ${\bf L}_{top}$ ranging between 0 and ${\bf M}_{top}$.  This is true in both the {\bf rep} and {\bf Verlinde bases}.   Thus we see that as $k$ increases the magic of the link states has a richer variety of long-range and residual subsystem magic.  We provide tables (table \ref{table:2linksk2} and table \ref{table:2linksk4}) of long-range mana for $k=2$ and $k=4$ link states in appendix \ref{app:linktable}, but for a handful of two-links we extend this investigation to $k=6$ link states\footnote{We propose the name {\it qusept} link states.} and chart the topological mana and the topological long-range mana in figure \ref{fig:2linkLRMbar}.  
\\
\\
Another feature of interest we find is that the topological long-range mana is always non-negative:
\beq
{\bf L}_{top}^{(\text{rep/Ver})}(k;\rho_{\mc L})\geq 0\qquad\qquad k=2,4\qquad \mc L\in\text{list}_{\text{two-links}}
\eeq
This motivates us to make the following conjecture:\\\\
{\bf Conjecture: }{\it All link states obey} super-additivity
\beq\label{eq:superadditivity}
\mc M_k(\rho)\geq\mc M_k(\rho_A)+\mc M_k(\rho_B)
\eeq
Although it is perhaps intuitive that a state should contain more magic than the sum of its subfactors, super-additivity is a stronger bound than subsystem concavity \eqref{eq:concavity}.  In fact it is possible to find states which are small deviations from product states on $\mc H_A\otimes \mc H_B$ that violate \eqref{eq:superadditivity} weakly\footnote{We thank Freek Witteveen for pointing this fact out to us and for the suggestion of using pseudo-random qutrit states to study long-range mana.}.  This is illustrated by the plots of figure \ref{fig:closetoprodplot}.
\\\\
The states violating \eqref{eq:superadditivity} are {\it very special}: it seems that typical states tend to obey superadditivity.  For instance, in figure \ref{fig:superadditivityplot} below, of 1000 pseudo-randomly generated states of two qutrits, {\it none} violated \eqref{eq:superadditivity}.  It would be very interesting to explore in more detail which states violate super-additivity and what role entanglement plays in either violating or protecting \eqref{eq:superadditivity}.  Indeed, some amount of entanglement is needed to violate \eqref{eq:superadditivity} (as $L_k=0$ for product states) however the plots in figure \ref{fig:closetoprodplot} suggest that enough entanglement can protect $L_k$ from being negative.  We suspect that none of states in figure \ref{fig:superadditivityplot} violate \eqref{eq:superadditivity} because typical states are almost never close to product states.  Thus we conjecture that the long-range entanglement of link-states (which is protected by topology) ensures that they satisfy super-additivity.  At this point, we leave the proof of this conjecture and general relation between entanglement and long-range mana as interesting open questions.
\begin{figure}[h!]
\centering
\includegraphics[width=.7\textwidth]{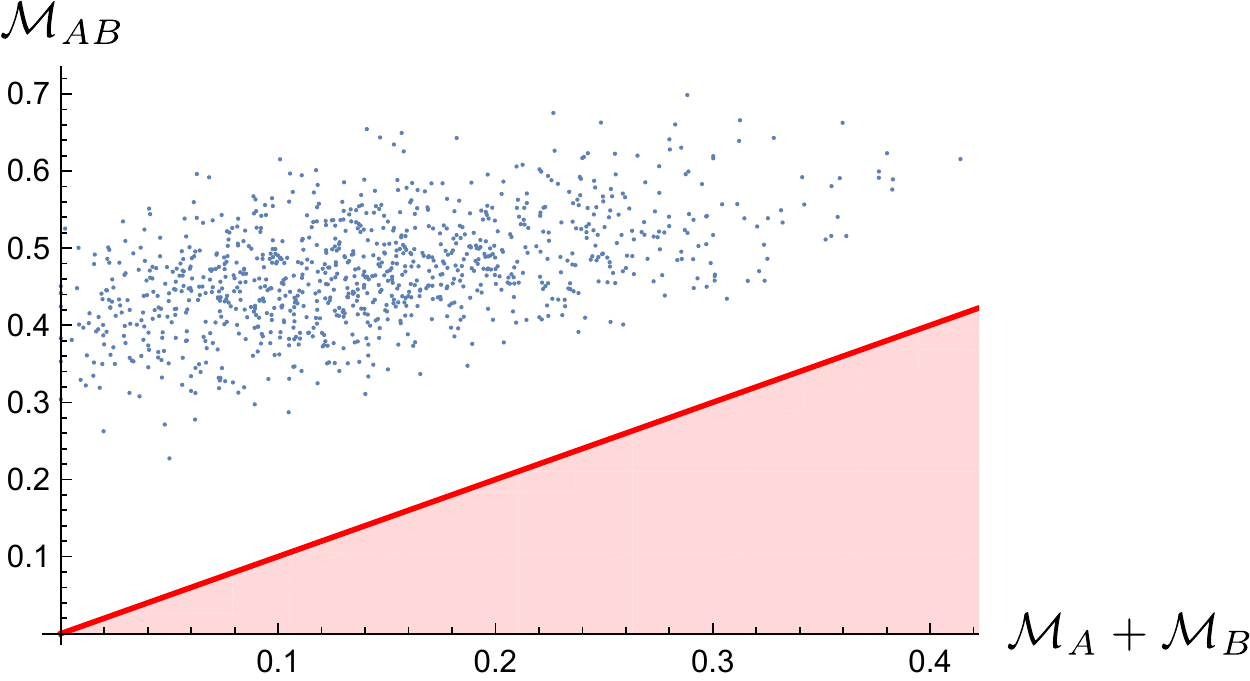}
\caption{\small{\textsf{A scatter plot of the global mana, $\mc M_{AB}\equiv \mc M(\rho)$, against the sum of subsystem mana, $\mc M_A+\mc M_B$, (where $\mc M_{A}=\mc M(\rho_{A})$ and $\mc M_B=\mc M(\rho_B)$) for 1000 pseudo-random states of two qutrits (i.e. $k=2$).  The region violating super-additivity, \eqref{eq:superadditivity}, is shaded in red.  The mixed states, $\rho$, are formed by starting with pure states on a three-qutrit Hilbert space, $|\psi\rangle_{p.r.}\in\mc H_3\otimes\mc H_3\otimes \mc H_3$, with pseudo-random complex coefficients (generated by Mathematica) and then tracing out the third qutrit.  $\rho_A$ ($\rho_B$) is subsequently formed from tracing out the second (first) qutrit}}}
\label{fig:superadditivityplot}
\end{figure}

\section{Discussion}\label{sect:disc}

In this paper we initiated a preliminary study into the magic of knot and link states in $SU(2)_k$ Chern-Simons theory.  The motivation for this was two-fold: (i) to quantify a notion of complexity to these states and (ii) to provide a simple yet controlled arena for exploring the long-range distribution of magic in quantum field theory.  The primary tool at our disposal is the state's mana, or the negativity of the its discrete Wigner function.  As a summary, we found that knot and link states are generically magical when prepared in a computation basis of Wilson loops of definite representation ({\bf rep basis}).  Because the wave-functions are constructed from the colored Jones polynomial, the magic of these states is extremely constrained at low values of $k$.  For knots, when the theory prepares qutrits ($k=2$) the nature of this magic is universal and entirely captured by the modular S-matrix.  This claim is bolstered by comparison to the states prepared in a basis diagonalizing the fusion rules ({\bf Verlinde basis}).  For ququint ($k=4$) and higher qudit theories, the amount of magic amongst knots is both more varied and more robust, often remaining non-zero in both computational bases.\\
\\
For links, we study the long-range magic by comparing the mana of the pure state with the mana of its subfactors.  We found that long-range magic is a generic feature of links states: almost every link has non-zero topological long-range mana, ${\bf L}_{top}$, and many possess a framing where the long-range mana saturates the global topological mana.  Although this is merely suggestive, for torus links in particular we are able to show that their reduced density matrices always have a framing that is stabilizer when prepared in the {\bf Verlinde basis,} and so their magic is always non-local.  In contrast to this, we also found many links whose reduced density matrices still possess residual long-range mana.  Thus while not entirely non-local, the magic of these states is more robust to reduction.  Lastly we conjectured that ${\bf L}_{top}$ is always positive for link-states, a property we name ``super-additivity," and which is not true of all quantum states.
\\
\\
There are several natural open questions resulting from this research and we discuss them below.\\
\\
{\bf Revisiting the framing}\\
\\
In this paper we have only focused on one feature of a knot or link's framing, which is the minimum of the mana over its orbit.  
The framing realizing this minimum, however, can differ from knot to knot and link to link.  An alternative approach would be to focus on a fixed framing\footnote{This could be fixed systematically through modifying the conventions of braid operators (as explained in appendix \ref{app:braidcomp}) in the construction of the wave-functions.}, distinguished by the minimal mana of a particularly simple knot or link.  This approach is not entirely free from ambiguity: for instance, while it might be appealing to compare knots in the fixed framing in which $\mc M^{(\text{Ver})}(\rho_{0_1})=0$, we found that for $k=2$ and $k=4$ there are actually two such framings.  For higher $k$, the minima could be more degenerate and how choose amongst them is unclear.  
As a separate issue, one might wonder, however, if there is any useful data contained in the {\it full list} of mana over the frame orbit.  This is large list of data (growing in $k$ as $(4k+8)^n$ for $n$-component links in $SU(2)_k$) and it is a shame to only keep one feature of it.  One potentially interesting feature, though counter to the interpretation of mana as lower bound, is the {\it maximum} mana over framings.  One reason is the following: for qutrit and ququint single party states it is known that mana is bounded above by (respectively)\cite{jain2020qutrit}
\beq
\mc M_{k=2}\leq\ln\left(\frac{5}{3}\right)\approx0.51\qquad\qquad\mc M_{k=4}\leq\sinh^{-1}\left(3+\sqrt 5\right)-\ln\left(5\right)\approx 0.748
\eeq
The pure states saturating these bounds are of interest because they are non-stabilizer Clifford eigenstates and candidates for the terminus of magic distillation \cite{Veitch:2014aa, van2011noise, jain2020qutrit,Prakash_2020}.  Of all the knots we considered in this paper, we did not find a single instance of framing saturating either bound.
 \\
\\
{\bf Diagnosing topologically insulating phases}\\
\\
The field theories investigated in this paper have a well-known connection to topological insulating phases, appearing as IR effective field theories.  The link invariants of these theories play a role in the characterization of the corresponding phase \cite{Wang_2020}. Of particular note in this paper is $SU(2)_2$ Chern-Simons theory which is relevant to the Moore-Read states \cite{Moore:1991ks} and related non-Abelian Pfaffian states \cite{Fradkin:1997ge}.  It is natural to question to what extent long-range magic provides a more {\it selective} diagnosis of topological order, signaling the presence of non-Abelian anyon statistics relevant for topological quantum computing.  
To some level the results of this paper show a ``proof-of-principle" that path integration in the effective field theory can reliably produce resource states for universal quantum computation, however since the topology of knot and link states in question are hard to prepare in any lab-ready systems, these comments are mostly speculative.  It is an open question if there is a suitable linear combination of magic monotones to unambiguously detect long-range magic (akin to the Kitaev-Preskill/Levin-Wen measure of long-range entanglement \cite{Kitaev:2005dm, PhysRevLett.96.110405}), in more physically relevant systems, however very recent work indicates that ``long-range magic"\footnote{There is some equivocation on terminology here: in particular, the definition of ``long-range magic" in \cite{ellison2020symmetryprotected} differs from the definition in this paper.} is a useful diagnostic of symmetry protected and many body phases \cite{ellison2020symmetryprotected, liu2020manybody}. Developing continuum effective field theory descriptions of these investigations is an interesting avenue for future research.\\
\\
{\bf $SL(2,\mathbb C)$ Chern-Simons and towards continuous Hilbert spaces.}\\
\\
Working in TQFT affords us several aspects of simplicity, namely finite dimensional Hilbert spaces and analytic control over the ground states, but it precludes some degree of applicability of our results to more generic quantum field theories.  In quantum mechanical systems with a continuum of states the question of ``stabilizer-ness" of a state is replaced by ``Gaussianity".  Here the (continuous) Wigner transform:
\beq
\mc W_{\vec p,\vec q}[\rho]=\frac{1}{(2\pi)^n}\int_{-\infty}^\infty d^n\vec u\,\langle \vec q-\vec u|\,\rho\,| \vec q+\vec u\rangle e^{i\vec p\cdot\vec u}
\eeq
again plays a central role.  For pure states, $\mc W$ is only positive (and thus a true joint probability distribution on phase space) iff $\rho$ is Gaussian \cite{hudson1974wigner,soto1983wigner}.  Thus the log-negativity of this function can serve as a signal of non-Gaussianity:
\beq\label{eq:mfM}
\mf M=\ln\int d^n\vec p\,d^n\vec q\,\big|\mc W_{\vec p,\vec q}[\rho]\big|.
\eeq
In our present context, one can imagine initiating a study of the negativity of continuum Wigner transforms in $SL(2,\mathbb C)$ Chern-Simons theory at large levels:
\beq
t=k+is,\quad \bar t=k-is\qquad\qquad k\rightarrow 0,\quad s=-i\sigma\rightarrow-i\infty
\eeq
The torus Hilbert spaces, in one choice of polarization, are spanned by the representation varieties of the meridian holonomy, $m=e^{u}$.  In the large $\sigma$ limit, this Hilbert space is spanned by a continuum of basis states and the path-integrals preparing the wavefunctions can be solved via saddle-point.  
For hyperbolic knots and links, there is always a dominating saddle at $\vec u=0$ called the {\it geometric branch,} and the expansion of the sub-leading corrections to this saddle are controlled by a $1/\sigma$ expansion \cite{Balasubramanian:2018por}:
\beq
\langle{\small\frac{1}{\sqrt\sigma}}\vec u|\mc L_{hyp}\rangle\sim\mc N\,e^{-\frac{\sigma}{\pi}V(\vec u)}\qquad\qquad V(\vec u)=V_{hyp}+\frac{1}{4\sigma}\sum_{i=1}^n\tau_{I,i}|u_i|^2+\ldots
\eeq
where $V_{hyp}$ is the hyperbolic volume of the link complement of $\mc L$.  {\it Thus at leading order in $1/\sigma$, all hyperbolic link state wave-functions are product states of Gaussian wave-packets and have vanishing $\mf M$.}  The $1/\sigma$ correction to the state is a quartic coupling in $V(\vec u)$\footnote{Also at this order is the first quantum correction to $\tau_{I,i}$, but this only affects the Gaussian term.} 
\beq
\left.V\right|_{\sigma^{-2}}=-\frac{1}{4\sigma^2}\sum_{ij}\text{Im}\left(u_i\overline{A_{ij}u_iu_j^2}-\frac{1}{2}A_{ij}u_i^2u_j^2\right)
\eeq
where $A_{ij}$ can be systematically computed from the gluing equations of the tetrahedral decomposition $\mc L$'s link complement (listed, for instance, in SnapPy \cite{SnapPy}).  To our knowledge there has not been any systematic study of these quartic coefficients in the vast mathematics literature involving hyperbolic 3-manifolds.  Yet from a physicist's perspective, it is interesting that they play a vital role in the structure of these states in more than one way: these same quartic coefficients are also the leading contribution to the entanglement structure \cite{Balasubramanian:2018por}.  It would be very interesting if continuum magic, perhaps in conjunction with multipartite entanglement, elucidated some yet-uncovered structure to these coefficients.  Unfortunately, it is easy to argue that while the first non-Gaussian contribution to the state is $\mc O(1/\sigma)$ the first non-zero contribution to $\mf M$ is actually $\mc O(1/\sigma^2)$\footnote{For instance, by supposition of a perturbative expansion in $\varepsilon=1/\sigma$ we assume that $\mf M$ is at least differentiable in $\varepsilon$ as $\varepsilon\rightarrow0$, however since it saturates its minimum at $\varepsilon=0$ the first derivative must vanish.  This can also just be verified by the definition of $\mf M$, \eqref{eq:mfM}, and keeping track of coefficients including the normalization of $|\mc{L}\rangle$.}, thus this problem suffers from the same subtleties as studying the entanglement entropy or entanglement negativity of these states, at least perturbatively.  Beyond mathematics, states in $SL(2,\mathbb C)$ Chern-Simons theory are of physical interest as well: in this limit the theory describes semi-classical Euclidean gravity with a negative cosmological constant \cite{witten19882+, Witten:1989ip, Gukov:2003na}.  It is an interesting question as to what extent $\mf M$ captures the transition of a distribution as a description of classical vs. quantum gravity.  This is not entirely unprecedented: in \cite{Balasubramanian:2005mg, Balasubramanian:2008da} it was noted that for holographic states of the 1/2-BPS sector of $\mc N=4$ super Yang-Mills, the non-negativity of $\mc W$ could be used as a necessary criteria for the emergence of a semi-classical bulk.  While hyperbolic link-states are not (conventionally) holographic, the vanishing of $\mf M$ at leading order dovetails nicely with this result.  Further connections between Wigner negativity and semi-classical bulk dynamics is an exciting avenue of ongoing research.
\\
\\
{\bf Revisiting ``complexity"}\\
\\
Lastly, while our notion of complexity is natural in the realm of quantum computation and QRT, from the viewpoint of the Chern-Simons path-integral (or the boundary tori hosting its Hilbert spaces) the classification of states based upon Pauli operators may seem contrived.  One might wonder if our program can be repeated with another (perhaps more natural) class of operators, in particular those generating the modular group $\{\mc S, \mc T\}$.  ``Circuit complexity" investigations along this line were made for torus knot states in \cite{Camilo:2019bbl} by counting the length of the minimum polynomial chain of $\mc S$ and $\mc T$'s needed to prepare the knot from the unknot.  Here we have in mind the complement of that program: that is, which knots cannot be prepared from $\mc S$ and $\mc T$ alone and how many non-modular operators are needed to prepare the state.  This definition of complexity, call it {\it modular knot magic}, has several appealing aspects: for one, it naturally solves the issue of framing ambiguity by fiat while still being sensitive to generic unitary transformations.  Additionally all torus knots, which can all be prepared from by polynomial strings of $\mc S$ and $\mc T$ acting on the unknot \cite{Camilo:2019bbl}, (and subsequently torus links) are considered ``simple" under this criterion.  Thus this modular knot magic would focus precisely on the knots and links whose classifications and colored Jones polynomials are more erratic.  This is perhaps as much a bug as a feature, but we regard it as an interesting open topic.

\vskip .5cm
{\bf Acknowledgements:} 
I would like to thank Jan de Boer, Michael Walter, Freek Witteveen, and Onkar Parrikar for helpful discussions.  I want to additionally thank Freek Witteveen and Jan de Boer for useful comments on the initial draft of this paper and an anonymous reviewer for comments towards the published version.  Lastly, I would like to thank the organizers of the June 2020 ``Complexity from Quantum Information to Black Holes" workshop and Brian Swingle for his presentation of \cite{White:2020zoz} and subsequent discussion from which the germ of this work originated. This work was supported by the ERC Starting Grant GenGeoHol. 

\appendix\numberwithin{equation}{section}

\section{Appendix: Computing colored Jones polynomials from braid closures}\label{app:braidcomp}
\begin{figure}[h!]
\centering
\begin{subfigure}[c]{.3\textwidth}
\includegraphics[width=3.5cm]{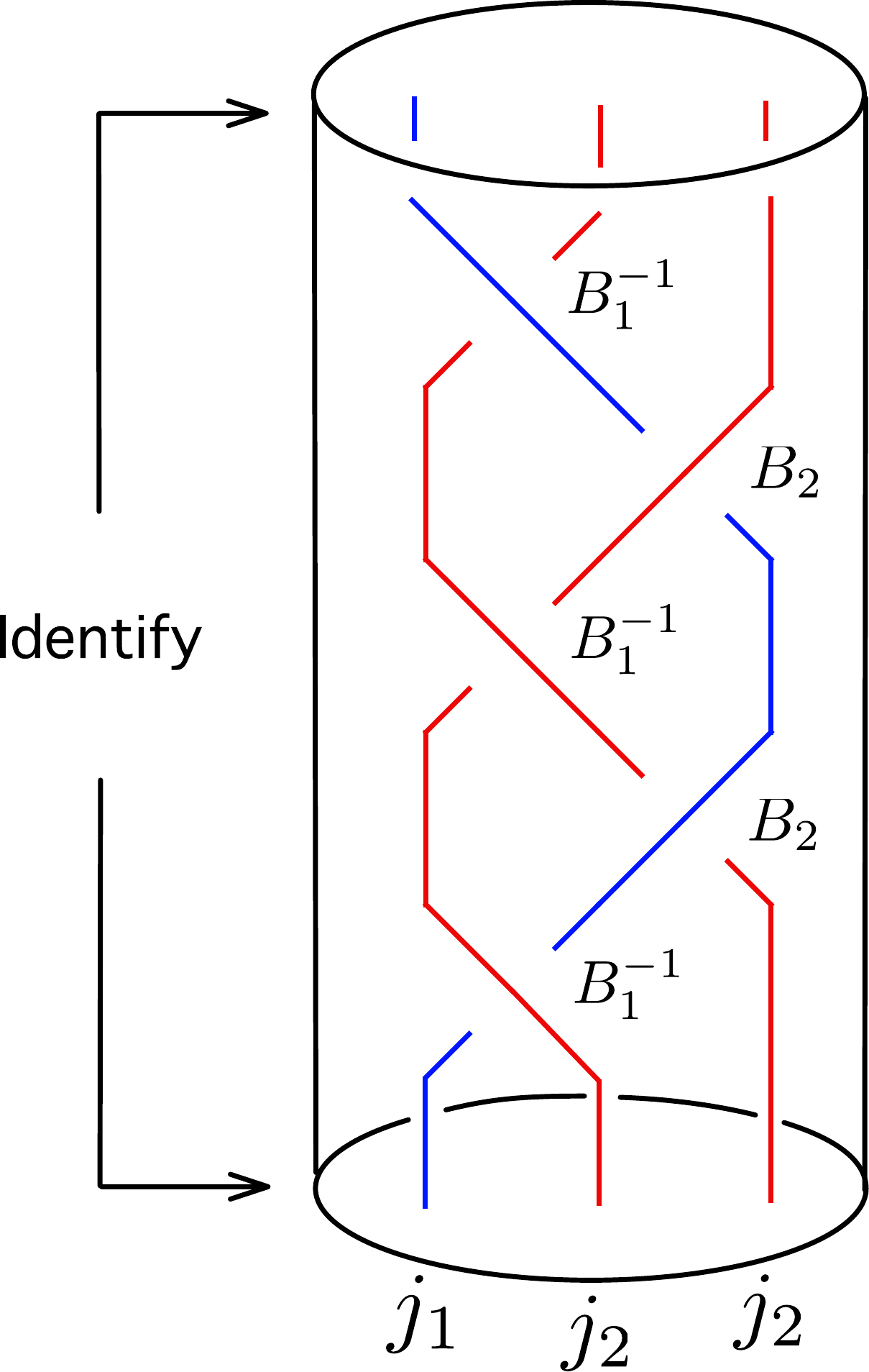}
\end{subfigure}
\begin{subfigure}[c]{.3\textwidth}
\includegraphics[width=4cm]{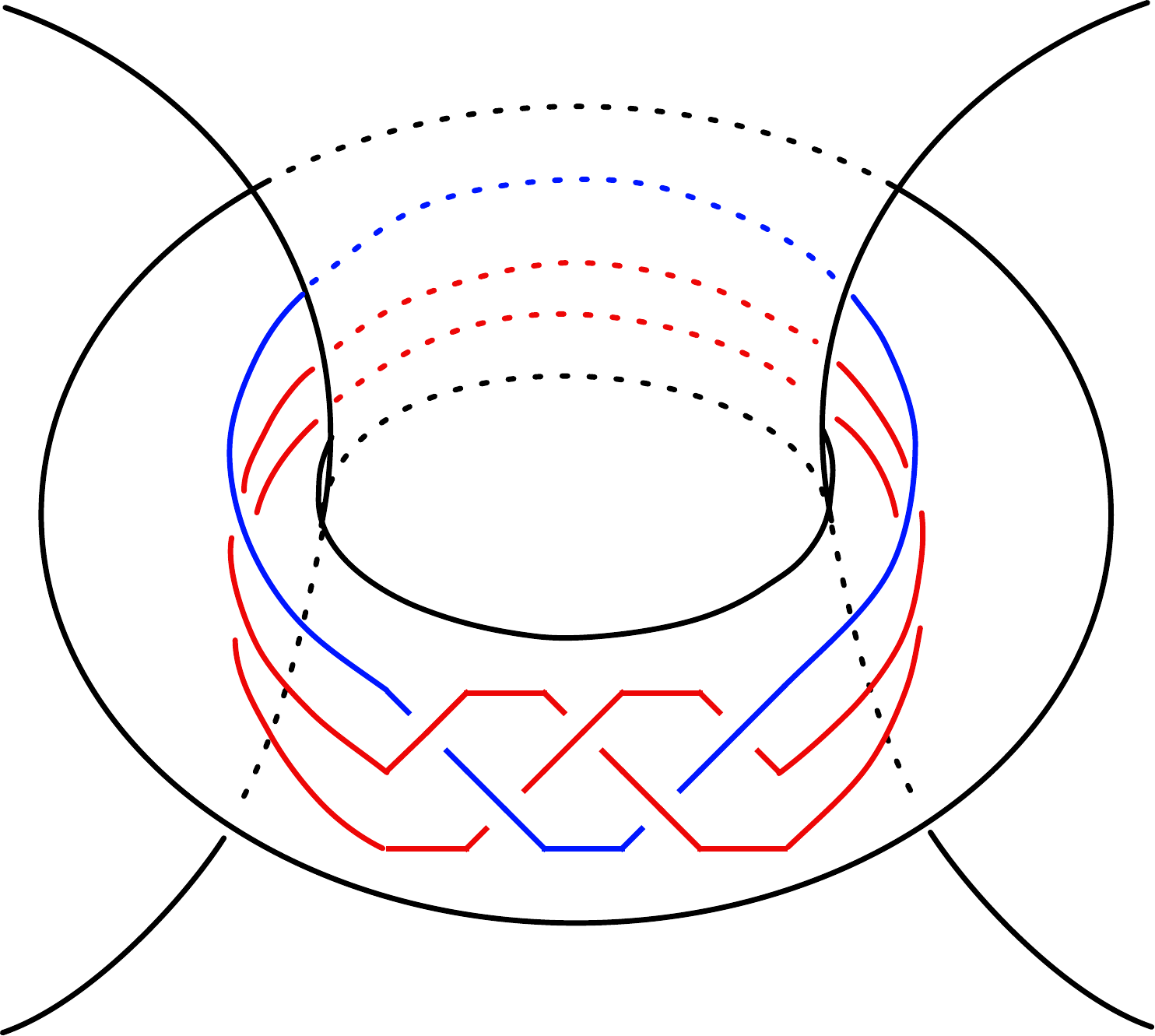}
\end{subfigure}
\begin{subfigure}[c]{.3\textwidth}
\includegraphics[width=6cm]{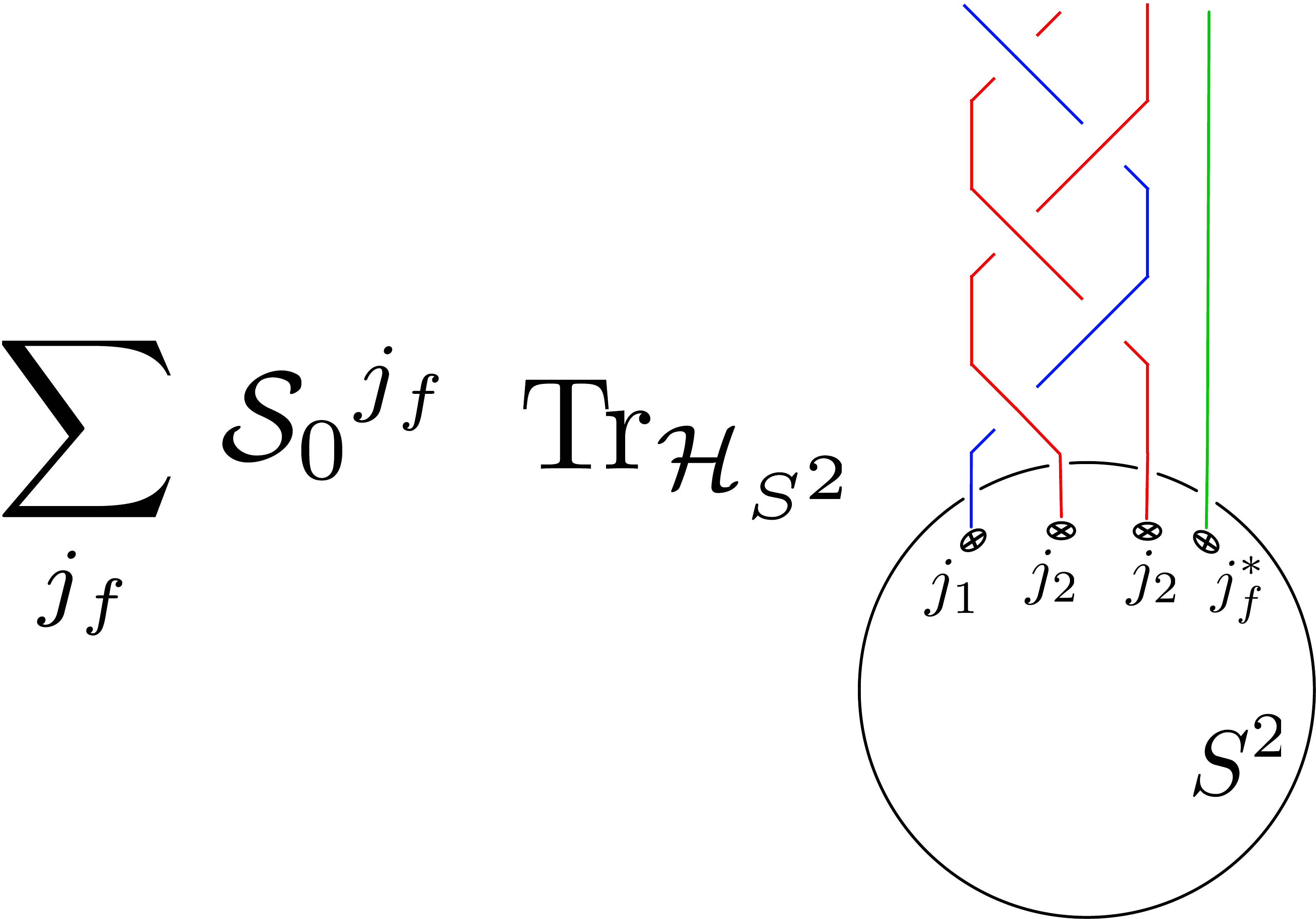}
\end{subfigure}
\caption{\small{\textsf{(Left) the braid closure of the Whitehead link (L5a1) is imbedded in a solid torus, $D^2\times S^1$.\\
(Middle) The remaining $S^3$ is filled with an (empty) solid torus identified along alternative cycles.\\
(Right) A modular S-transformation on the empty torus results in the path-integral on $S^2\times S^1$, or a trace over the punctured $S^2$ Hilbert space. The auxiliary strand, $j_f^\ast$, book-keeps the S-matrix elements.}}}
\label{fig:WHBR}
\end{figure}
\begin{figure}[h!]
\centering
\begin{subfigure}[b]{.4\textwidth}
\centering
\includegraphics[height=2.5cm]{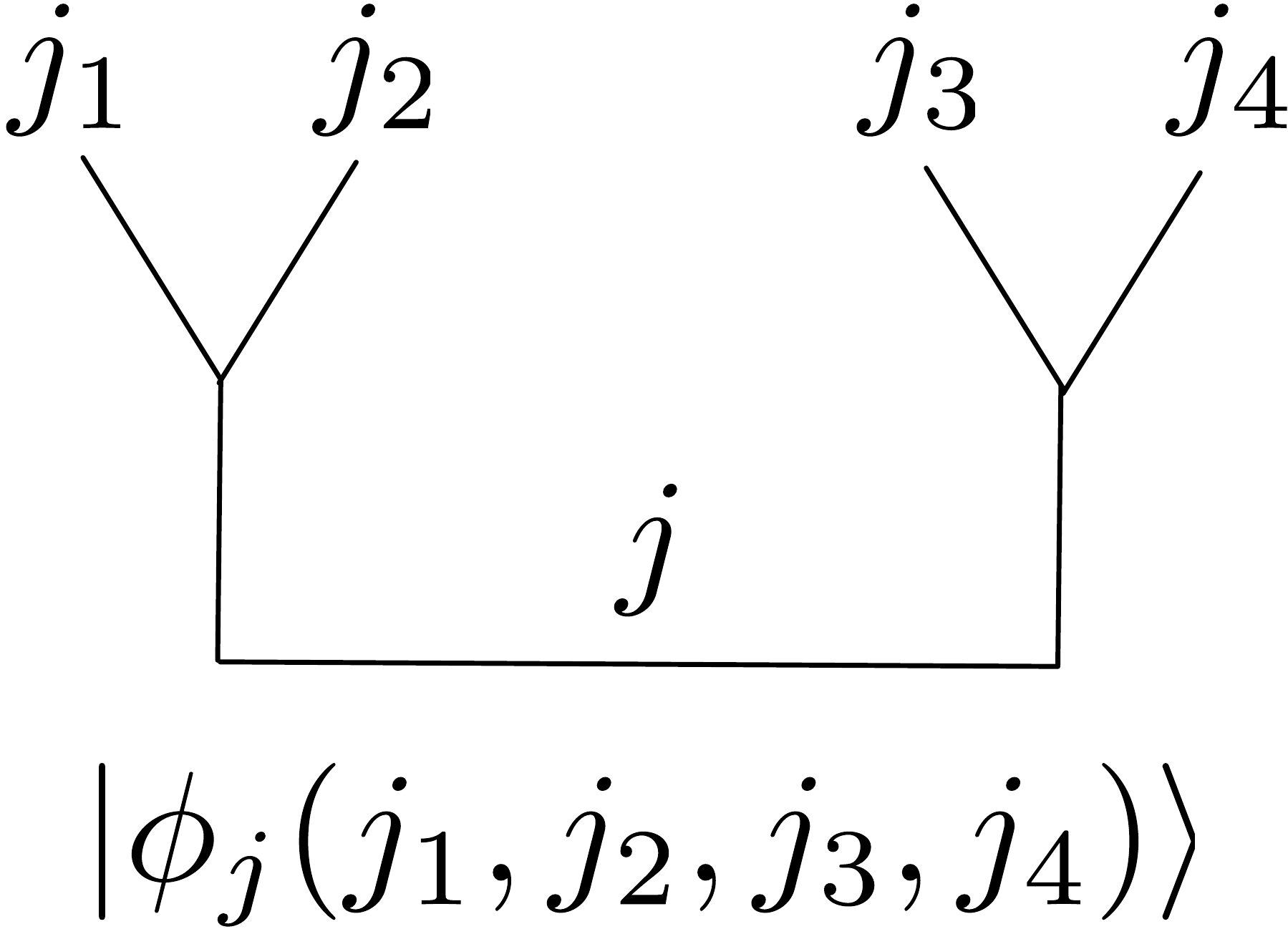}
\end{subfigure}
\begin{subfigure}[b]{.4\textwidth}
\centering
\includegraphics[height=2.5cm]{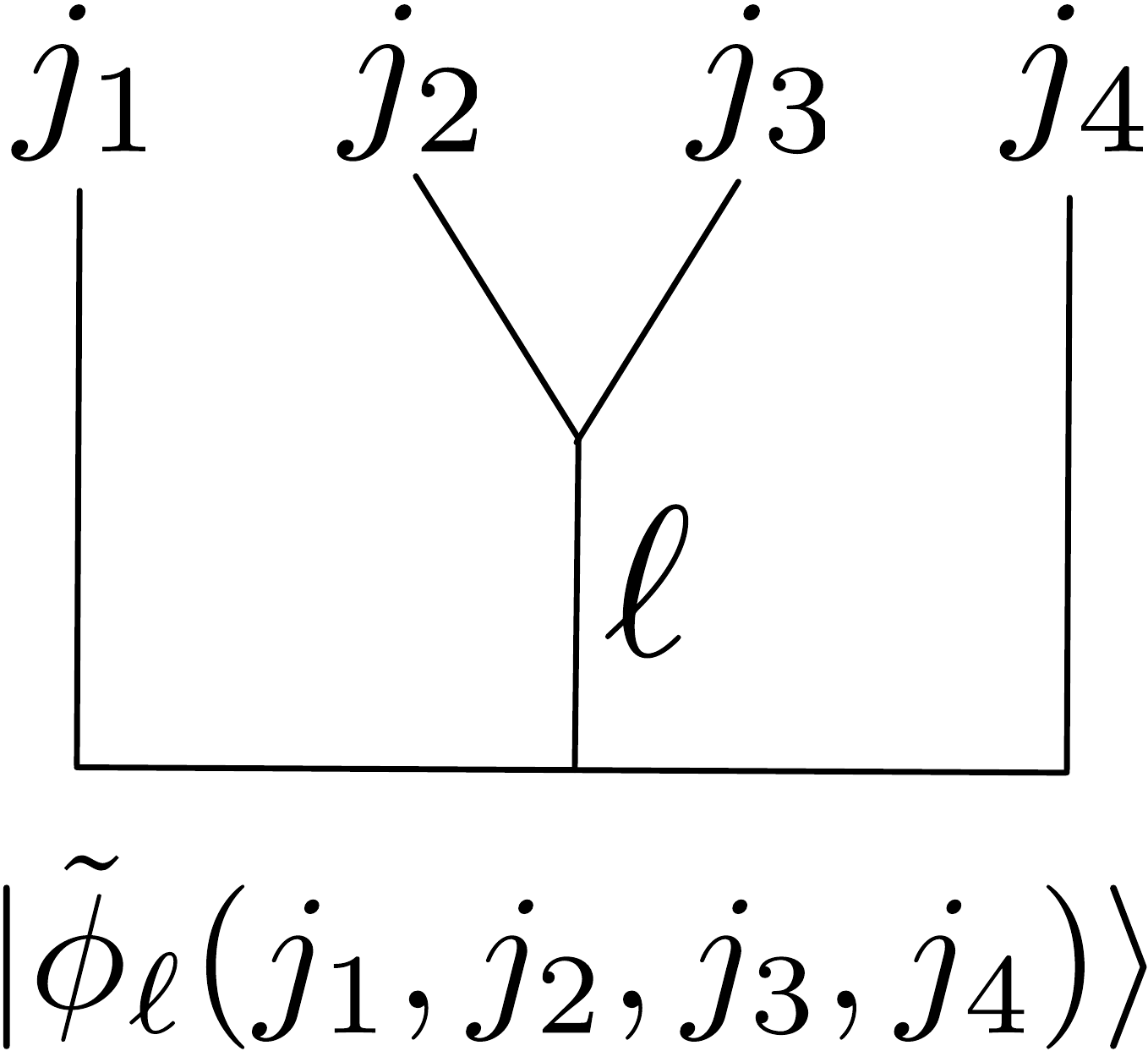}
\end{subfigure}
\caption{\small{\textsf{Two different bases for the $SU(2)_k$ 4-point conformal blocks.}}}
\label{fig:CBbases}
\end{figure}

In this appendix we review our method for computing the wave-functions of knot and link states in $SU(2)_k$ Chern-Simons theory.  Given the presentation of a link, $\mc L$, as the closure of a braid, $\mc B_{\mc L}$, with $n_b$ strands (note that $n_b$ might be larger than the number of components of $\mc L$) we imagine embedding  $\mc B_{\mc L}$ inside a solid torus $D^2\times S^1$ with the ``time" of the braid running along the $S^1$.  This solid torus is then embedded inside $S^3$ by gluing an empty $D^2\times S^1$ along the alternative cycles.  Implementing a modular S-transformation on this empty solid torus and gluing it back onto our braid-filled torus results in a path-integral on $S^2\times S^1$.  In doing so we need to keep track of the modular S-matrix elements.  Inserting the identity $\sum_{j_f}|j_f\rangle\langle j_f|$ on $\mc H_{T^2}$ picks out the ${\mc S_0}^{j_f}$ component, however inserts an auxiliary ``book-keeping" Wilson loop running along the $S^1$.  The resulting path-integral is a trace of an extended braid operator acting on the $n_b+1$ punctured $S^2$:
\beq\label{eq:tmcBdef}
\langle \prod_{i=1}^nW_{j_i}(L_i)\rangle=\sum_{j_f}{\mc S_0}^{j_f}\,\mTr_{\mc H_{S^2(j_1,\ldots,j_{n_b},j_f^\ast)}}\left(\tmc B_{\mc L\cup 0_1}(j_1,\ldots,j_{n_b},j_f^\ast)\right)
\eeq
where $\tmc B_{\mc L\cup 0_1}$ is the braid-closure obtained from $\mc B$ by adding a disconnected unknot.  The $\ast$ on $j_f^\ast$ comes from a flip in orientation.  This whole procedure is illustrated as a cartoon in figure \ref{fig:WHBR}.\\
\\
It is well-known that this Hilbert space is spanned by the ${n_b}+1$ point conformal blocks of the $SU(2)_k$ WZW conformal field theory \cite{Elitzur:1989nr} (see \cite{Moore:1988qv} for an extensive review of background material).
We will evaluate the trace in the $|\phi_{\bs j}(j_1,j_2,\ldots, j_{n_b+1})\rangle$ basis (pictured for $n_b+1=4$ on the left-hand side of figure \ref{fig:CBbases}) which are eigenvectors of the {\it odd} braid operators
\beq
\hat B_{2i-1}|\phi_{\bs j}(j_1,\ldots, j_{n_b+1})\rangle=\lambda_{j_{2i-1,2i}}(j_{2i-1},j_{2i})|\phi_{\bs j}(j_{2i-1}\leftrightarrow j_{2i})\rangle
\eeq
with eigenvalue\footnote{Typically the eigenvalues $\lambda_j$ come equipped with a $\pm$ superscript to distinguish the relative orientation of the strands being braided.  Because the links are realized as braid-closures, we can always orient the strands in the same direction along the $S^1$ and so this detail will not be pertinent for us.}
\beq\label{eq:braidev}
\lambda_{j_{ab}}(j_a,j_b)=(-1)^{j_a+j_b-j_{ab}}q^{\frac{j_a(j_{a}+1)}{2}+\frac{j_{b}(j_{b}+1)}{2}-\frac{j_{ab}(j_{ab}+1)}{2}}\qquad\qquad q=e^{\frac{2\pi i}{k+2}}.
\eeq
Note that encoded in these eigenvalues is a {\it particular choice} of framing (in particular our choice here differs from that in \cite{Kaul:1993hb,Kaul:1991np}).  Although there is no ``preferred" choice of framing, in \ref{sect:manarev} we will account for this ambiguity by varying the results over all framings at the end.
\\\\
Eigenvectors of the even braid operators are the alternative conformal blocks $|\tilde\phi_{\bs \ell}(j_1,\ldots, j_{n_b+1})\rangle$, depicted for $n_b+1=4$ on the right-hand side of figure \ref{fig:CBbases}:
\beq
\hat B_{2i}|\tilde\phi_{\bs\ell}(j_1,\ldots, j_{n_b+1})\rangle=\lambda_{\ell_{2i,2i+1}}(j_{2i},j_{2i+1})|\tilde\phi_{\bs \ell}(j_{2i}\leftrightarrow j_{2i+1})\rangle
\eeq
with $\lambda_\ell$ given by the same formula \eqref{eq:braidev}.  Focusing on $n_b=3$, the two bases are related by 4-point fusion duality coefficients\footnote{To be more precise, $a_{j\ell}\left(\begin{array}{cc}j_1&j_2\\j_3&j_4\end{array}\right)$ are defined as above if $j$ and $\ell$ are allowed by the fusion algebra:
\begin{align}
&\text{max}\left(|j_1-j_2|,|j_3-j_4|\right)\leq j\leq \text{min}\left(j_1+j_2,k-j_1-j_2,j_3+j_4,k-j_3-j_4\right)\nonumber\\
&\text{max}\left(|j_1-j_4|,|j_2-j_3|\right)\leq \ell\leq \text{min}\left(j_1+j_4,k-j_1-j_4,j_2+j_3,k-j_2-j_3\right)
\end{align}
otherwise they are defined to be zero.}
\begin{align}
a_{j\ell}\left(\begin{array}{cc}j_1&j_2\\j_3&j_3\end{array}\right)\equiv&\langle\phi_j(j_1,j_2,j_3,j_4)|\tilde\phi_\ell(j_1,j_2,j_3,j_4)\rangle\nonumber\\
=&
(-1)^{j_1+j_2+j_3+j_4}\sqrt{[2j+1][2\ell+1]}\left(\begin{array}{ccc}j_1&j_2&j\\j_3&j_4&\ell\end{array}\right)
\end{align}
determined by the $q$-Racah coefficients
\begin{align}
\left(\begin{array}{ccc}j_1&j_2&j\\ j_3&j_4&\ell\end{array}\right)=&\Delta(j_1,j_2,j)\Delta(j_3,j_4,j)\Delta(j_1,j_4,\ell)\Delta(j_2,j_3,\ell)\nonumber\\
&\sum_{m\geq 0}(-1)^m[m+1]!\big\{[m-j_1-j_2-j]![m-j_3-j_4-j]![m-j_1-j_4-\ell]!\nonumber\\
&\;\;[m-j_2-j_3-\ell]![j_1+j_2+j_3+j_4-m]![j_1+j_3+j+\ell-m]![j_2+j_4+j+\ell-m]!\big\}^{-1}
\end{align}
where
\beq
\Delta(j_1,j_2,j_3)=\sqrt{\frac{[-j_1+j_2+j_3]![j_1-j_2+j_3]![j_1+j_2-j_3]!}{[j_1+j_2+j_3+1]!}}
\eeq
and
\beq
[x]=\frac{q^{x/2}-q^{-x/2}}{q^{1/2}-q^{-1/2}},\qquad [x]!=\prod_{i=1}^x[i].
\eeq
Higher point duality coefficients can be built out of the successive applications of the 4-point coefficients, although we will not need them for this paper.\\
\\
To see how this works out practically, let us take the Whitehead link, L5a1.  It can be represented as the closure of
\beq
\mc B_{\text{L5a1}}=B_1^{-1}B_2B_1^{-1}B_2B_1^{-1}.
\eeq
acting on three-strands (the last two must be in the same representation), as pictured in the left-subfigure of figure \ref{fig:WHBR}.  Adding in the book-keeping strand, we arrive at the following colored Jones polynomial (up to some overall normalization)
\begin{align}
\langle j_1,j_2|\text{L5a1}\rangle={\mc N}^{-1/2}\sum_{j_f}\sum_{p_1,p_2}\sum_{q_1,q_2}&{\mc S_0}^{j_f}\lambda_{p_1}^{-1}(j_1,j_2)\lambda_{q_1}(j_1,j_2)\lambda_{p_2}^{-1}(j_2,j_2)\lambda_{q_2}(j_2,j_1)\nonumber\\
&a_{p_1q_1}\left(\begin{array}{cc}j_2&j_1\\j_2&j_f\end{array}\right)a_{p_2q_1}\left(\begin{array}{cc}j_2&j_2\\j_1&j_f\end{array}\right)a_{p_2q_2}\left(\begin{array}{cc}j_2&j_2\\j_1&j_f\end{array}\right)a_{p_1q_2}\left(\begin{array}{cc}j_2&j_1\\j_2&j_f\end{array}\right)
\end{align}
where the sum on $j_f$ ranges over all values, $0,\,1/2,\,\ldots,\,k/2$ and the sums on $p_1,\,p_2,\,q_1,$ and $q_2$ range over the values allowed by the fusion algebra.  One can verify that this result agrees with other representations of the Whitehead colored Jones polynomial (e.g. \cite{Habiro0}).

\section{Appendix: Table of topological mana for knots}\label{app:listknots}
The set of knots considered in this paper are knots with 10 crossings or less that have a braid-closure representative in the KnotTheory Mathematica package \cite{KnotTheory} containing three strands or less.  The full list of knots is
{\footnotesize\begin{align}
\text{list}_{\text{knots}}=\{&0_1,\,3_1,\,4_1,\,5_1,\,5_2,\,6_2,\,6_3,\,7_1,\,7_3,\,7_5,\,8_2,\,8_5,\,8_7,\,8_9,\,8_{10},\,8_{16},\,8_{17},\,8_{18},\,8_{19},\,8_{20},\,8_{21},\,9_1,\,9_3,\,9_6,\nonumber\\
&\,9_9,\,9_{16},\,10_2,\,10_5,\,10_9,\,10_{17},\,10_{46},\,10_{47},\,10_{48},\,10_{62},\,10_{64},\,10_{79},\,10_{82},\,10_{85},\,10_{91},\,10_{94},\,10_{99},\nonumber\\
&\,10_{100},\,10_{104},\,10_{106},\,10_{109},\,10_{112},\,10_{116},\,10_{118},\,10_{123},\,10_{124},\,10_{125},\,10_{126},\,10_{127},\,10_{139},\,10_{141},\nonumber\\
&\,10_{143},\,10_{148},\,10_{149},\,10_{152},\,10_{155},\,10_{157},\,10_{159},\,10_{161}\}
\end{align}}
All of these are in ``Rolfsen notation," $C_n$ where $C$ lists the number of crossings and the subscript $n$ is the knot's position in the Rolfsen table.\\
\\
Below we list the minimum mana computed, ${\bf M}_{top}$, the knot states in both the {\bf rep computational basis} and in the {\bf Verlinde computational basis}.  For the qutrit theory ($k=2$) all knots in $\text{list}_{\text{knots}}$ had identical topological mana, depending on the computational basis:
\beq
{\bf M}_{top}^{(\text{rep})}(2;\rho_{\mc K})=\ln\left(\frac{1}{3}(1+2\sqrt 2)\right)\approx.243842\qquad\qquad {\bf M}_{top}^{(\text{Ver.})}(2;\rho_{\mc K})=0.
\eeq
For $k=4$ the ququint topological mana in both computational bases is tabulated below:
{\footnotesize
\begin{longtable}{ |p{1.5cm}||p{2.5cm}|p{2.5cm}|  }
 \hline
 $\mc K$ & ${\bf M}_{top}^{(\text{rep})}(4;\rho_{\mc K})$ & ${\bf M}_{top}^{(\text{Ver.})}(4;\rho_{\mc K})$\\
 \hline
$0_1$&0.39677&0\\
$3_1$&0.581537&0.28036\\
$4_1$&0.39677&0\\
$5_1$&0.39677&0\\
$5_2$&0.39677&0\\
$6_2$&0.39677&0\\
$6_3$&0.39677&0\\
$7_1$&0.39677&0\\
$7_3$&0.39677&0\\
$7_5$&0.39677&0\\
$8_2$&0.39677&0\\
$8_5$&0.581537&0.28036\\
$8_7$&0.39677&0\\
$8_9$&0.39677&0\\
$8_{10}$&.0581537&0.28036\\
$8_{16}$&0.39677&0\\
$8_{17}$&0.39677&0\\
$8_{18}$&0.620695&0.417243\\
$8_{19}$&0.581537&0.28036\\
$8_{20}$&0.581537&0.28036\\
$8_{21}$&0.581537&0.28036\\
$9_{1}$&0.581537&0.28036\\
$9_{3}$&0.39677&0\\
$9_{6}$&0.581537&0.28036\\
$9_{9}$&0.39677&0\\
$9_{16}$&0.581537&0.28036\\
$10_{2}$&0.39677&0\\
$10_{5}$&0.581537&0.28036\\
$10_{9}$&0.581537&0.28036\\
$10_{17}$&0.39677&0\\
$10_{46}$&0.39677&0\\
$10_{47}$&0.39677&0\\
$10_{48}$&0.39677&0\\
$10_{62}$&0.581537&0.28036\\
$10_{64}$&0.581537&0.28036\\
$10_{79}$&0.39677&0\\
$10_{82}$&0.581537&0.28036\\
$10_{85}$&0.581537&0.28036\\
$10_{91}$&0.39677&0\\
$10_{94}$&0.39677&0\\
$10_{99}$&0.620695&0.417243\\
$10_{100}$&0.39677&0\\
$10_{104}$&0.39677&0\\
$10_{106}$&0.581537&0.28036\\
$10_{109}$&0.39677&0\\
$10_{112}$&0.581537&0.28036\\
$10_{116}$&0.39677&0\\
$10_{118}$&0.39677&0\\
$10_{123}$&0.39677&0\\
$10_{124}$&0.39677&0\\
$10_{125}$&0.39677&0\\
$10_{126}$&0.39677&0\\
$10_{127}$&0.39677&0\\
$10_{139}$&0.581537&0.28036\\
$10_{141}$&0.581537&0.28036\\
$10_{143}$&0.581537&0.28036\\
$10_{148}$&0.39677&0\\
$10_{149}$&0.39677&0\\
$10_{152}$&0.39677&0\\
$10_{155}$&0.39677&0\\
$10_{157}$&0.39677&0\\
$10_{159}$&0.581537&0.28036\\
$10_{161}$&0.39677&0\\
 \hline
\end{longtable}}

\section{Appendix: Tables of topological long-range mana for links}\label{app:linktable}

The two-links considered in this paper are those with 10 crossings or less whose braid-closure representatives in KnotTheory possess three strands or less.  There are 34 such links in total.  In ``Hoste-Thistlethwaite notation"\footnote{In the expression ``L9a31", for instance, ``L" stands for ``link" (as opposed to a ``K" for knot), ``9" lists the number of crossings, ``a" stands for ``alternating" (as opposed to ``n" for ``non-alternating") and ``31" is the bibliographical listing of that link amongst the other alternating links with 9 crossings.} the complete list is
{\footnotesize\begin{align}
\text{list}_{\text{two-links}}=\{&\text{L2a1,  L5a1,  L7a1,  L7a3,  L7a6,  L7n1,  L7n2,  L8a14,  L9a2,  L9a9,  L9a14,  L9a20,}\nonumber\\
&\text{L9a21,  L9a22,  L9a28,  L9a29,  L9a31,  L9a36,  L9a38,  L9a39,  L9a41,  L9a42,  L9n4,}\nonumber\\
&\text{L9n5,  L9n6,  L9n13,  L9n14,  L9n15,  L9n16,  L9n17,  L9n18,  L9n19,  L10a118}\}
\end{align}}
The topological mana and topological long-range mana in both the {\bf rep} and the {\bf Verlinde} computational bases for the states in $\text{list}_{\text{two-links}}$ are tabulated in table \ref{table:2linksk2} for the qutrit theory $(k=2)$ and in table \ref{table:2linksk4} for the ququint theory $(k=4)$.
{\footnotesize
\begin{longtable}{ |p{1.5cm}||p{2.5cm}|p{2.5cm}||p{2.5cm}|p{2.5cm} | }
\hline
 $\mc L$ & ${\bf M}_{top}^{(\text{rep})}(2;\rho_{\mc L})$ & ${\bf L}_{top}^{(\text{rep})}(2;\rho_{\mc L})$ & ${\bf M}_{top}^{(\text{Ver.})}(2;\rho_{\mc L})$ & ${\bf L}_{top}^{(\text{Ver.})}(2;\rho_{\mc L})$\\
 \hline
 L2a1&0.69098&0.69098&0.398092&0.398092\\
 L5a1&0.838387&0.838387&0.455746&0.455746\\
 L6a3&0.69098&0.69098&0.398092&0.398092\\
 L7a1&0.838387&0.838387&0.455746&0.455746\\
 L7a3&0.487684&0&0&0\\
 L7a6&0.69098&0.69098&0.398092&0.398092\\
 L7n1&0.756718&0.365926&0.342347&0.278952\\
 L7n2&0.838387&0.838387&0.455746&0.455746\\
 L8a14&0.838387&0.838387&0.455746&0.455746\\
 L9a2&0.838387&0.838387&0.455746&0.455746\\
 L9a9&0.487684&0&0&0\\
 L9a14&0838387&0.838387&0.455746&0.455746\\
 L9a20&0.69098&0.69098&0.398093&0.398092\\
L9a21&0.69098&0.69098&0.398093&0.398092\\
L9a22&0.69098&0.69098&0.398092&0.398092\\
L9a28&0.69098&0.69098&0.398092&0.398092\\
L9a29&0.69098&0.69098&0.398092&0.398092\\
L9a31&0.69098&0.69098&0.398092&0.398092\\
L9a36&0.756718&0.365926&0.342347&0.278952\\
L9a38&0.487684&0&0&0\\
L9a39&0.756718&0.365926&0.342347&0.278952\\
L9a41&0.756718&0.365926&0.342347&0.278952\\
L9a42&0.838387&0.838387&0.455746&0.455746\\
L9n4&0.756718&0.365926&0.342347&0.278952\\
L9n5&0.487684&0&0&0\\
L9n6&0.838387&0.838387&0.455746&0.455746\\
L9n13&0.69098&0.69098&0.398092&0.398092\\
L9n14&0.69098&0.69098&0.398092&0.398092\\
L9n15&0.69098&0.69098&0.398092&0.398092\\
L9n16&0.69098&0.69098&0.398092&0.398092\\
L9n17&0.69098&0.69098&0.398092&0.398092\\
L9n18&0.487684&0&0&0\\
L9n19&0.487684&0&0&0\\
L10a118&0.69098&0.69098&0.398092&0.398092\\
 \hline
 \caption{\small{\textsf{The $k=2$ global topological mana and the topological long-range mana for a set of two-links in both the {\bf rep} and {\bf Verlinde bases.}}}}
 \label{table:2linksk2}
\end{longtable}}

{\footnotesize
\begin{longtable}{ |p{1.5cm}||p{2.5cm}|p{2.5cm}||p{2.5cm}|p{2.5cm} | }
\hline
 $\mc L$ & ${\bf M}_{top}^{(\text{rep})}(4;\rho_{\mc L})$ & ${\bf L}_{top}^{(\text{rep})}(4;\rho_{\mc L,red})$ & ${\bf M}_{top}^{(\text{Ver.})}(4;\rho_{\mc L})$ & ${\bf L}_{top}^{(\text{Ver.})}(4;\rho_{\mc L,red})$\\
 \hline
 L2a1&1.32224&1.32224&0.640231&0.640231\\
 L5a1&1.36052&0.818751&1.25679&0.530504\\
 L6a3&1.31841&1.31841&0.641391&0.641391\\
 L7a1&1.30665&0.52305&1.05058&0.283174\\
 L7a3&1.30418&0.77074&1.14547&0.528229\\
 L7a6&1.2522&1.2522&0.589438&0.589438\\
 L7n1&1.2823&0.482255&1.1085&0.28442\\
 L7n2&1.30418&0.777074&1.14547&0.528229\\
 L8a14&1.304143&1.304143&0.695988&0.695988\\
 L9a2&1.30418&0.777074&1.14547&0.528229\\
 L9a9&1.30665&0.523052&1.05058&0.283174\\
 L9a14&0.79354&0&0&0\\
 L9a20&1.29623&1.15973&1.22391&1.11277\\
L9a21&1.32224&1.32224&0.640231&0.640231\\
L9a22&1.32224&1.32224&0.640231&0.640231\\
L9a28&1.36321&1.25239&1.1455&0.76138\\
L9a29&1.32587&1.21473&1.23202&1.1082\\
L9a31&1.3456&1.24135&1.16145&0.932038\\
L9a36&1.27081&0.7714&1.18748&0.419378\\
L9a38&1.26875&0.315405&0.828224&0.215018\\
L9a39&1.30143&1.30143&0.695988&0.695988\\
L9a41&1.2449&0.140127&0.741035&0.0988102\\
L9a42&1.28969&0.892507&0.932162&0.396929\\
L9n4&1.19829&0.47611&0.495705&0.291176\\
L9n5&1.36052&0.818751&1.25679&0.530504\\
L9n6&1.30665&0.523052&1.05058&0.283174\\
L9n13&1.3333&0.752818&1.17563&0.641731\\
L9n14&1.32587&1.21473&1.23202&1.1082\\
L9n15&1.36321&1.25239&1.1455&0.76138\\
L9n16&1.36321&1.25239&1.1455&0.76138\\
L9n17&1.32587&1.21473&1.23202&1.1082\\
L9n18&1.2449&0.140127&0.741035&0.0988102\\
L9n19&1.281&0.42431&0.951281&0.242791\\
L10a118&1.2522&1.2522&0.589438&0.589438\\
 \hline
 \caption{\small{\textsf{The $k=4$ topological global mana and the topological long-range mana for a set of two-links in both the {\bf rep} and {\bf Verlinde bases.}}}}
 \label{table:2linksk4}
\end{longtable}}

\providecommand{\href}[2]{#2}\begingroup\raggedright\endgroup

\end{document}